\documentclass[manuscript]{acmart}

\AtBeginDocument{%
  }

\setcopyright{acmlicensed}
\copyrightyear{2027}
\acmYear{2027}
\acmDOI{XXXXXXX.XXXXXXX}

\acmConference[Conference acronym 'XX]{Make sure to enter the correct
  conference title from your rights confirmation email}{June 03--05,
  2018}{Woodstock, NY}
\acmISBN{978-1-4503-XXXX-X/18/06}

\begin{document}

\title{From LLM Hallucinations to Inspiration in Creative Work}

\author{Xinrui Lin}
\email{x.lin@bit.edu.cn}
\orcid{0000-0003-3632-5563}
\affiliation{%
  \institution{Beijing Institute of Technology}
  \city{Beijing}
  \country{China}}

\author{Heyan Huang}
\authornote{Corresponding author.}
\email{hhy63@bit.edu.cn}
\orcid{0000-0002-0320-7520}
\affiliation{%
  \institution{Beijing Institute of Technology}
  \city{Beijing}
  \country{China}}

\author{Shuning Zhang}
\email{zsn23@mails.tsinghua.edu.cn}
\orcid{0000-0002-4145-117X}
\affiliation{%
  \institution{Tsinghua University}
  \city{Beijing}
  \country{China}}

\author{Qi Wang}
\email{qi.wang@bit.edu.cn}
\orcid{0000-0002-5728-4062}
\affiliation{%
  \institution{Beijing Institute of Technology}
  \city{Beijing}
  \country{China}}

\author{Youchao Zhou}
\email{yczhou@bit.edu.cn}
\orcid{0009-0005-3563-5625}
\affiliation{%
  \institution{Beijing Institute of Technology}
  \city{Beijing}
  \country{China}}

\author{John Vines}
\authornotemark[1]
\email{john.vines@ed.ac.uk}
\orcid{0000-0003-4051-3356}
\affiliation{%
  \institution{University of Edinburgh}
  \city{Edinburgh}
  \country{United Kingdom}}

\renewcommand{\shortauthors}{Lin et al.}

\begin{abstract}
Large language models (LLMs) can introduce unexpected ideas through hallucinations, yet little is known about when creative practitioners find these errors inspiring or how these judgments shape interactions. We conducted a retrospective qualitative study with 20 experienced practitioners in research, design, and social media content creation, thematically analyzing 41 hallucination cases and 20 semi-structured interviews. Our framework links inspiration judgments of hallucinations based on goal relevance, task stage, content adaptability, and type-specific factors to behaviors across not-inspired, inspired-by, and inspired-to states. We characterize downstream interaction consequences, showing how hallucination-inspired experiences vary across creative stages and reshape role-differentiated trust, and propose these consequences as a dimension of hallucination evaluation. Finally, we derive design recommendations for stage-sensitive control of divergence and adherence, user-defined constraint priorities, differentiated hallucination labels, and salvage support for preserving useful ideas while correcting errors. These contributions deepen HCI understanding of hallucinations and responsible LLM design in creative work.

\end{abstract}



\received{12 September 2026}
\received[revised]{10 December 2026}
\received[accepted]{16 January 2027}

\maketitle

\section{Introduction}
Large language models (LLMs), such as ChatGPT, are increasingly used in creative practice to generate, organize, and develop ideas. Yet these systems can produce outputs that appear plausible while departing from facts, user instructions, or the conversational context. Such outputs are commonly described as hallucinations, although their definitions vary across studies \cite{huang_survey_2025, rawte_troubling_2023, ji_survey_2023, ye_cognitive_2023, li_lokis_2025, zhang_sirens_2025}. Following Huang et al., we distinguish factual hallucinations, which conflict with verifiable real-world facts, from faithfulness hallucinations, which depart from user instructions, the given context, or the output's own logic \cite{huang_survey_2025}. Examples include fabricated references \cite{rawte_troubling_2023} and responses that ignore stated requirements \cite{huang_survey_2025}. Hallucinations have been linked to limited access to current information \cite{ray_chatgpt_2023} and to limitations in model reasoning and memory \cite{mckenna_sources_2023}. This phenomenon has even been argued to be an inevitable, inherent limitation of LLMs \cite{xu_hallucination_2025}.

Hallucinations are generally unwelcome and can cause severe harm when decisions depend on accurate information. Documented harms include fabricated legal precedents \cite{dahl_large_2024}, unsafe medical advice \cite{pal_med-halt_2023}, and inappropriate reliance on generated outputs \cite{spatharioti_effects_2025}. These failures can also change how users assess and trust LLMs in human-computer interaction (HCI) \cite{lin_towards_2024, yin_lies_2024}. Prior research has consequently focused on evaluating, detecting, and mitigating hallucinations \cite{tonmoy_comprehensive_2024}. Evaluation and detection approaches include benchmarks such as HaluEval \cite{li_halueval_2023}, semantic-uncertainty measures \cite{farquhar_detecting_2024}, and user-facing explanations, verification cues, and comparisons of repeated outputs \cite{reinhard_fact_2025, ehsan_new_2025, cheng_relic_2024, leiser_hill_2024}. Model-level mitigation includes retrieval-augmented generation \cite{lewis_retrieval-augmented_2020, shi_hallucination_2023, agrawal_can_2024}, consistency-based selection \cite{wang_self-consistency_2023}, self-verification \cite{dhuliawala_chain--verification_2024}, and knowledge editing \cite{meng_locating_2022, meng_mass-editing_2023}. This work is essential for reducing unreliable outputs, but it offers less insight into what users do with hallucinations that still arise in LLM use.

In creative work, however, the consequences of hallucinations may be more varied. Hallucinations have been described as a form of "computational imagination" because unexpected departures can introduce associations that users might not otherwise consider \cite{pilcher_purposefully_2025, lan_schizosyntax_2026}. Some studies suggest both a positive association between hallucination and creativity and a trade-off between reducing hallucinations and preserving novelty or divergent generation \cite{lee_mathematical_2023, he_shakespearean_2025, jiang_survey_2024, lan_schizosyntax_2026}. Qualitative research further shows that users may treat some erroneous outputs as material for ideation \cite{holmstrom_what_2026}. By departing from factual grounding or user requirements, these outputs can break with conventional logic and introduce unexpected associations. As the beginning of creative work, inspiration may benefit from such outputs because it involves being moved by a creative idea beyond one's existing understanding and becoming motivated to act on it \cite{thrash_inspiration_2003, thrash_inspiration_2004, thrash_psychology_2014}. Prior work in writing and artistic practice has begun to examine hallucinations as a potential source of inspiration \cite{halperin_artificial_2024, wan_it_2024}. Research on LLM-supported inspiration and creativity also suggests that inaccurate outputs can open new directions, but this potential varies across contexts \cite{lin_seeking_2025, holmstrom_what_2026}. For instance, in science fiction writing, hallucinations may broaden the imaginative space while also introducing misleading scientific errors \cite{argese_hallucination_2026}. Thus, hallucinations remain errors, but whether they inspire creative practitioners, and how that inspiration shapes creative work, cannot be inferred from incorrectness alone.

Taken together, prior work establishes that hallucinations can be harmful, that they can sometimes provide material for creativity. It remains unclear, however, under what circumstances creative practitioners working with LLMs judge a particular factual or faithfulness hallucination to be inspiring. We also know little about how this judgment shapes their subsequent behavior, cognition, and interaction experiences, or how LLMs might preserve useful ideas without encouraging unreliable generation. We therefore ask:

\begin{description}
    \item[\textbf{RQ1:}] What factors shape whether creative practitioners perceive an LLM hallucination as inspiring?
    \item[\textbf{RQ2:}] How do creative practitioners respond after judging an LLM hallucination as inspiring or not inspiring?
    \item[\textbf{RQ3:}] How do these judgments shape creative practitioners' cognition and interaction experiences with LLMs?
    \item[\textbf{RQ4:}] What do creative practitioners expect from LLMs in leveraging the creative potential of hallucinations while mitigating their risks?
\end{description}

To address these research questions, we conducted a retrospective qualitative study involving 20 participants with extensive experience using LLMs in research, design and social media content creation. During one-on-one sessions, participants reviewed hallucinations in their provided chat logs and color-coded whether each case evoked a new possibility and motivated further action. We then conducted semi-structured interviews to understand their judgments, subsequent actions, interaction experiences, and design views. We performed a thematic analysis of 41 cases alongside 20 interview transcripts. Our findings suggest that judging a hallucination as inspiring is a process in which users assess whether its unexpected content exceeds their prior knowledge and remains relevant to their goals, given their task stage and circumstances. This study makes following contributions to HCI:
\begin{enumerate}
    \item We present a framework connecting creative practitioners' situated judgments of hallucinations with subsequent behaviors across the not-inspired, \emph{inspired-by}, and \emph{inspired-to} states. When hallucinations transcend users' knowledge, these judgments depend on goal relevance, task stage, and whether the errors can be adapted for use. We also identify factors specific to inspiration from factual and faithfulness hallucinations.
    \item We characterize downstream interaction consequences, showing how hallucination-inspired experiences vary across creative stages and reshape role-differentiated trust, and propose using these consequences as one dimension for evaluating hallucinations.
    \item We derive LLM design recommendations for stage-sensitive control over divergence and adherence, user-defined constraint priorities, differentiated hallucination labels, and \emph{salvage support} that supports inspiration while correcting errors.
\end{enumerate}

\section{Related Work}
\subsection{Inspiration}

Inspiration has been studied in social psychology for approximately 40 years. It is primarily defined as an intrinsic motivational state triggered by external sources and is connected to the realization of new ideas, which is associated with creative expression \cite{bottger_customer_2017, stephan_nostalgia-evoked_2015, thrash_inspiration_2003, baird_inspired_2012}. Our understanding of inspiration mainly based on the research of Thrash et al.
They defined inspiration as a motivational state that impels an individual to transform newly acquired ideas into action \cite{thrash_inspiration_2010, thrash_mediating_2010, thrash_psychology_2014}. Specifically, inspiration must be evoked by an external stimulus or object; it symbolizes cognitive transcendence - the discovery of better possibilities previously unknown - and motivates the individual to realize these new possibilities \cite{thrash_inspiration_2003}. Therefore, the core characteristics of inspiration are evocation (being evoked by stimuli), transcendence (transcending the everyday or mundane), and approach motivation (expressing or actualizing these new ideas) \cite{thrash_inspiration_2003}. 

Core to this understanding of inspiration lies in describing the process by which these characteristics manifest as the juxtaposition of two component processes: the “inspired-by” state and the “inspired-to” state \cite{bottger_customer_2017, thrash_chapter_2021, thrash_inspiration_2004}. The “inspired-by” state is the cognitive phase of recognizing and appreciating better possibilities in the evoked object that one had never imagined, while the “inspired-to” state represents the motivational phase involving the extension and implementation of the qualities embodied by the evoked object \cite{thrash_inspiration_2004}. This theory also emphasizes that inspiration does not occur when people merely appreciate external objects without taking any subsequent action \cite{thrash_inspiration_2004}.
Therefore, in this paper, we use "inspired-by" to explain participants' expressions of appreciation for LLM hallucination that go beyond their prior knowledge or awareness, while "inspired-to" is used to indicate where they expressed motivations to actualize the idea.

\subsection{Hallucination in HCI}


In this paper, following Huang et al.'s taxonomy \cite{huang_survey_2025}, we distinguish between factual and faithfulness hallucinations. Although they depart from different references, both types involve generated content that is inconsistent with a reference relevant to the current task. This distinction allows us to examine how users verify external facts and judge whether outputs align with their intentions and constraints. Hallucination evaluation, automatic detection, and model-level mitigation have formed a substantial body of research \cite{huang_survey_2025, li_halueval_2023, farquhar_detecting_2024, tonmoy_comprehensive_2024}. This section instead focuses on how HCI research understands users' judgments of hallucinations and intervenes to support their responses.


HCI research has mainly used warnings, explanations, sources, and verification cues to help users identify and handle factual hallucinations. These designs can improve recognition, but may also increase cognitive load, encourage reliance on surface cues \cite{nahar_fakes_2024, reinhard_fact_2025, tian_when_2026, kim_fostering_2025}. For outputs that depart from requirements, HCI research has also helped users express constraints on format, length, semantics, and style \cite{liu_we_2024}. Existing work further shows that users may rely on incorrect answers, even incorporating them into final outputs when they conflict with their own correct answers. Such reliance and trust vary with the form of explanation, prior experience, expertise, and task risk \cite{kim_fostering_2025, spatharioti_effects_2025, ashktorab_emerging_2025, ryser_calibrated_2025}. For clearly abnormal or nonsensical outputs, users' prior experience and attribution of intent also shape how they experience them. Users may interpret such outputs as system errors, anthropomorphize the system and feel uneasy, or accept them as part of a game's narrative and mechanics \cite{rapp_how_2025, yin_lies_2024}. Overall, these studies show that the effects of hallucinations depend on the user and task context. However, prior work has primarily examined factual errors and their effects on detection, decision-making, reliance, and trust. Existing work has made limited comparisons between factual and faithfulness hallucinations. Understanding the role of hallucinations for inspiration therefore requires connecting these findings with research on how LLMs support inspiration and creativity.

\subsection{LLM for Inspiration and Creativity}


Using LLMs for creativity and ideation has been an important topic in HCI. Research suggests that, on some creativity measures, LLM-generated ideas can be more original than those produced by people with lower creativity scores \cite{domanti_are_2026}. In design \cite{sattele_creative_2025, chen_autospark_2024}, research \cite{pu_ideasynth_2025}, writing \cite{chakrabarty_creativity_2024}, and education \cite{fan_litlinker_2025}, LLMs have helped users ideate, organize materials, and develop ideas. Similar support has been explored in script-writing \cite{tang_how_2026}, media production \cite{choi_proxona_2025}, and visual expression \cite{marquardt_imaginationvellum_2025, wang_aideation_2025}. LLMs can also use cross-domain analogies to support creative thinking during problem definition, ideation, implementation, and evaluation \cite{li_beyond_2026}. Against this background, LLM hallucinations are no longer viewed solely as harmful defects. Research, especially in creative writing and artistic expression, has begun to consider them as potential sources of creativity \cite{lee_mathematical_2023, jiang_survey_2024, halperin_artificial_2024, bangerl_creaitive_2025, wan_it_2024, pilcher_purposefully_2025, he_shakespearean_2025, argese_hallucination_2026, lan_schizosyntax_2026}. Unexpected content from hallucinations may encourage divergent thinking and help users move beyond established patterns of thought. Accordingly, hallucinations have also been described as a form of computational imagination \cite{pilcher_purposefully_2025, lee_mathematical_2023, he_shakespearean_2025, jiang_survey_2024}. As a key component of creative expression \cite{thrash_mediating_2010}, inspiration often benefits from this and has already been shown to arise from LLM-generated ideas and analogies that depart from their references \cite{lin_seeking_2025, kang_biospark_2024}.


However, whether users can draw inspiration from hallucinations and accept their creative value may depend on the context and requires human judgment \cite{lan_schizosyntax_2026}. In education, a qualitative study of master's students found that hallucinations could expand divergent thinking but also hinder creative work through over-reliance \cite{holmstrom_what_2026}. In open-ended domains such as art and writing, hallucinations may be treated as unexpected content worth exploring, but pose more direct risks in domains that prioritize factual accuracy, such as law and healthcare \cite{chakraborty_promethean_2024, dahl_large_2024, pal_med-halt_2023}. Hallucinations can be double-edged even within the same creative task. For example, they may expand the imaginative space of science fiction while introducing misleading scientific inaccuracies \cite{argese_hallucination_2026}. Overall, prior work shows that hallucinations can be a source of inspiration. However, it has mostly examined whether LLMs generate novel content, support ideation, or offer creative value in particular contexts. We still lack systematic research grounded in the real-world practices to explain which factors shape whether creative practitioners are inspired by hallucinations, and examine how this inspiration affects their subsequent behavior, cognition, and interaction experiences with LLMs.

\section{Methods}

Prior work suggests that hallucinations, LLM outputs departing from facts or user requirements, are more likely to provide inspiration in creative domains \cite{holmstrom_what_2026, lin_seeking_2025, kang_biospark_2024}. We therefore studied experienced LLM users' real creative tasks to examine how they negotiate facts, constraints, and possibilities, and when unexpected outputs become worth pursuing.

We conducted a retrospective qualitative study anchored in participants' LLM chat logs. Recruitment prioritized users who had been inspired by a hallucination and retained the relevant logs. We made this choice for three reasons. First, participants' concrete tasks and subsequent actions supported detailed comparisons among meaningless hallucinations, merely interesting hallucinations and those that prompted exploration. The logs also provided a record for checking these accounts. Second, a controlled study could not reliably produce particular hallucinations, while deliberately inducing them for study materials could mislead participants. Third, shared logs grounded interviews in specific contexts, helping participants recall when they noticed a hallucination, what followed, and how later turns changed their judgment. This grounding reduced reliance on general recollections. All submitted logs predated the study and came from participants' everyday activities. Participants chose their own platforms, and we analyzed only volunteered logs without requiring registration for or use of any designated service. The data included logs from ChatGPT, DeepSeek, and Doubao.

We recruited experienced practitioners in research, design, and social media content creation. These domains respectively produce knowledge, products and visual designs, and public-facing content. All three are representative forms of creative work requiring continual idea generation, assessment, and development. In research, factual hallucinations such as fabricated papers may suggest questions. Faithfulness hallucinations such as incomplete or scope-omitting answers may prompt reframing or reveal overlooked directions. In design, fabricated products, data, or features may suggest categories and attributes. Departures from style, structure, or material constraints may yield unexpected prototype concepts or interaction approaches. In social media content creation, incorrect information or browsing data may prompt new topics and authentic sources during verification. Departures from plot or setting requirements may reveal narrative gaps or suggest new forms of expression and plot directions. Together, these domains cover distinct outputs and situations in which both hallucination types may inspire. The university's institutional ethics review board approved the study.

\subsection{User Study Procedure}\label{sm:procedure}

Each participant completed a one-on-one session conducted by the first author. The researcher explained our definitions of factual and faithfulness hallucinations \cite{huang_survey_2025} and Thrash and Elliot's theory of inspiration \cite{thrash_inspiration_2003, thrash_inspiration_2004}. The researcher checked participants' understanding before continuing. Participants brought existing LLM chat logs containing at least two hallucinations, including at least one they had already considered inspiring. The researcher and participant jointly classified each case. They verified factual hallucinations online and identified faithfulness hallucinations by checking departures from prompts, context, or internal logic. After validating the cases, participants completed a color-coding task and a semi-structured interview. Sessions lasted approximately 70 minutes.

\subsubsection{color-coding}\label{sm:color-coding}

Participants shared log excerpts in Microsoft Word and could remove private or irrelevant content before submission. Each excerpt retained the relevant context, prompts, and hallucinations. Using Word's collaborative editing, the researcher and participant reviewed each case in context and coded the hallucination green, yellow, or red. Green indicated a new possibility that prompted participants to explore, verify, thus experiencing \emph{inspired-by} and \emph{inspired-to} states. Yellow indicated novel or interesting content that the participant did not pursue, remaining in the \emph{inspired-by} state. Red indicated unacceptable or meaningless content that did not reach the \emph{inspired-by} state. For example, a fabricated paper was green if its abstract suggested a research idea that the participant explored in later prompts. It was yellow if the participant found it novel but unrelated to the current goal and did not intend to pursue it.

For each highlighted hallucination, the researcher asked: 
\begin{enumerate}
    \item Why the participant considered it a hallucination and whether it was factual or faithfulness?
    \item Why the participant coded that color and, for green cases, what the hallucination contributed, such as a concrete outcome or change to an idea?
    \item What the participant did next, how those actions affected inspiration, and, for green cases, how the hallucination became a feasible idea?
    \item Whether other log content, hallucinations, or subsequent actions affected the color result?
    \item Hypothetical recoding: whether changed goals, constraints, or context could change the color. For example, a yellow case that prompted no action in the current task could become green by prompting exploration elsewhere, or red by losing its meaning.
\end{enumerate}
The first four questions reconstructed the participant's original judgment and subsequent actions. The fifth examined how goals, constraints, and task stages might change the same hallucination's color. The researcher recorded each response in the Word document. Participants confirmed that its conversational context, color, and explanation remained aligned.

\subsubsection{Semi-structured Interview}\label{sm:interview}

After color-coding, we conducted a semi-structured interview about experiences beyond the submitted cases and comparisons across hallucination types and stages of creative work. With consent, we audio-recorded and transcribed the interviews. Participants rated their LLM experience in creative work from 1, no familiarity, to 5, expert-level experience (Section \ref{sm:participants}). They need to discuss how often hallucinations inspired them, their general views, harms caused by hallucinations, and how they avoided those harms. Then participants identified characteristics of inspiring and non-inspiring hallucinations and described how inspiration from each type varied across workflow stages of creative work. We also examined how other hallucinations in a long conversation affected current judgments and how hallucination-inspired experiences differed from suggestions that followed facts and requirements.

The interview then examined what participants did after making different judgments. For factual hallucinations, we first asked participants how they identified them. If they suspected first, we inquired whether inspiration preceded verification, whether participants verified first, and whether discovering an error during verification could still surface inspiring material. If they had confirmed it is a hallucination, we asked whether it inspired participants and what they did after either judgment. For faithfulness hallucinations, we examined behaviors after inspiring and non-inspiring cases and how tighter constraints affected their occurrence and inspirational potential. We also asked how hallucination-inspired experiences shaped ethical judgments, trust, and the interaction experience. Participants considered whether hallucinations should be eliminated or retained and how LLMs could use their inspirational potential while avoiding harm. The researcher followed up on submitted cases throughout. Participants could discuss the two types separately or combine them when the distinction was difficult to apply. 

\subsection{Participants}\label{sm:participants}

\begin{table*}[t]
  \centering
  \caption{Participant demographics, creative work domains, self-rated LLM experience, and submitted hallucination cases ("Fact." denotes factual hallucination, "Faith." denotes faithfulness hallucination, and "--" indicates that no case was submitted for that color).}
  \label{tab:participants}
  \small
  \begin{tabular}{ccclcccc}
    \toprule
    ID & Age & Gender & Domain & LLM Exp. & Green & Yellow & Red \\
    \midrule
    P1  & 26 & Male   & Research     & 4.0       & 1 Fact.  & --        & 1 Faith. \\
    P2  & 30 & Female & Design       & 3.5       & 1 Fact.  & 1 Fact.   & 1 Fact. \\
    P3  & 24 & Male   & Design       & 4.0       & 1 Faith. & --        & 1 Fact. \\
    P4  & 25 & Female & Research     & 4.0       & 1 Fact.  & 1 Fact.   & -- \\
    P5  & 27 & Male   & Research     & 4.5       & 2 Faith. & --        & -- \\
    P6  & 25 & Female & Social Media & 4.0       & 1 Fact.  & --        & 1 Faith. \\
    P7  & 24 & Male   & Research     & 3.5       & 1 Faith. & --        & 1 Fact. \\
    P8  & 23 & Female & Design       & 4.0       & 1 Fact.  & --        & 1 Fact. \\
    P9  & 27 & Male   & Social Media & 4.0       & 1 Faith. & 1 Faith.  & -- \\
    P10 & 20 & Female & Design       & 4.0       & 1 Fact.  & 1 Fact.   & -- \\
    P11 & 27 & Male   & Research     & 3.5       & 1 Faith. & --        & 1 Faith. \\
    P12 & 25 & Male   & Social Media & 4.0       & 1 Fact.  & 1 Faith.  & -- \\
    P13 & 24 & Male   & Design       & 4.0       & 1 Faith. & 1 Faith.  & -- \\
    P14 & 32 & Female & Social Media & 5.0       & 1 Fact.  & --        & 1 Fact. \\
    P15 & 25 & Male   & Design       & 4.5       & 1 Faith. & --        & 1 Faith. \\
    P16 & 33 & Male   & Social Media & 4.0       & 1 Faith. & 1 Faith.  & -- \\
    P17 & 26 & Female & Social Media & 4.0       & 1 Faith. & --        & 1 Faith. \\
    P18 & 24 & Female & Social Media & 4.5       & 1 Faith. & --        & 1 Fact.+Faith. \\
    P19 & 37 & Female & Social Media & 4.5       & 1 Fact.  & --        & 1 Fact. \\
    P20 & 26 & Male   & Design       & 4.5       & 1 Fact.  & --        & 1 Fact. \\
    \bottomrule
  \end{tabular}
\end{table*}

We recruited 20 participants through university email lists and social media and identified them as P1--P20. Each received RMB 70 and completed the full procedure. Table \ref{tab:participants} reports demographics, self-rated LLM experience, and submitted cases by color. The sample included 11 men and 9 women: 5 in research, 7 in design, and 8 in social media content creation. Social media practices included creative food videos (P6), sleep-aid videos (ASMR; P9), knowledge videos (P12), travel recommendations (P14), script-writing for short videos or role-playing social games (P16--18) and art-education videos (P19). Participants were 20--37 years old ($M=26.50$, $SD=3.87$). All participants rated their LLM experience at 3.5 or above ($M=4.10$, $SD=0.38$), indicating that they considered themselves experienced LLM users.

Participants submitted 41 color-coded hallucinations: 21 green, 7 yellow, and 13 red. Twenty were factual, 20 were faithfulness, and one belonged to both categories. P5 submitted two similar green faithfulness hallucinations. We retained both because they appeared at different log positions and the participant confirmed each separately. P18's red case both contradicted external facts and violated prompt constraints, so Table \ref{tab:participants} marks it as both a factual and a faithfulness hallucination.

\subsection{Data Analysis}\label{sm:data-analysis}

The data comprised 20 Word documents, one per participant, containing complete submitted LLM chat logs and color-coding results, plus 20 semi-structured interview transcripts. The first author analyzed them using Braun and Clarke's thematic analysis \cite{braun_using_2006}. To become familiar with the corpus, the first author repeatedly read all materials and linked each participant's logs, color-coding results, and interview responses by ID. During transcript coding, the first author consulted the corresponding logs to identify the context and subsequent actions associated with each judgment.

During initial coding, the first author coded every statement relevant to the RQs and organized codes into broad RQ-aligned categories. RQ1 distinguished hallucination type, factual or faithfulness, with reasons for inspiration or no inspiration, such as "RQ1--Factual hallucinations: reasons for inspiration". RQ2 used the same cross-classification for behaviors following inspiration or no inspiration, like "RQ2--Faithfulness hallucinations: behaviors following inspiration". RQ3 covered "Comparison with normal suggestions", "Faithfulness hallucinations: perceived inspiration under changing constraints", and "Trust and relationships". RQ4 covered "User control", "Hallucination identification, labeling, and mitigation", and "Extracting inspiration from hallucinations". Each category contained more specific codes. For example, if a participant, when verifying a fabricated paper, finds real and relevant research and is then inspired, we applied the codes "RQ1--Factual hallucination: reasons for inspiration--Relevant to the goal", "RQ1--Factual hallucination: reasons for inspiration--Easy to verify", and "RQ2--Factual hallucination: behavior following inspiration--Verify first, then become inspired". Statements with similar meanings across participants and domains received the same code. Each excerpt retained the participant ID, creative work domain, hallucination type, and source, either the color-coding results or interview.

In the next round, we compared positive and negative cases and grouped related codes into candidate themes, arranged candidate themes into four RQ-aligned groups. For example, green cases were coded "Relevant to the goal" and yellow or red cases were coded "Departing from the current goal" together informed the theme "RQ1--Shared across hallucination types--Goal relevance is necessary for inspiration". We reviewed candidate themes against all 41 cases and 20 transcripts, comparing commonalities and differences across colors and the three creative-work domains. Participants' hypothetical recoding during color-coding helped test theme boundaries. When a changed goal or constraint altered the same case's proposed color, we revised interpretations that attributed the result to content alone. Using these boundary checks, we defined and named themes through several rounds of merging and splitting while retaining their RQ1--RQ4 organization. For the Findings, we selected cases, interview excerpts, and quotations that clearly illustrated each theme. Participant counts report theme distributions within this dataset, not population prevalence.

\section{Findings}

In reporting our findings, we identify participants as P1--P20 and treat novelty beyond their prior knowledge as a prerequisite for hallucination-inspired experiences because transcendence is central to inspiration. The Findings therefore examine when novel departures from facts or user requirements moved participants from the \emph{inspired-by} to the \emph{inspired-to} state. When discussing the color-coding results, we use the definitions introduced in the Methods section: green cases prompted exploration or action, yellow cases were novel or interesting but did not motivate action, and red cases were unacceptable or meaningless. We specify the hallucination type and creative context where relevant.

\subsection{Factors Shaping Whether Hallucinations Inspire}\label{sf:factors}

This subsection addresses RQ1 by identifying the shared and type-specific factors that shaped whether participants were inspired by hallucinations  (Table \ref{tab:inspiration-factors}). Across both types, inspiration required novelty beyond prior knowledge and goal relevance, and was more likely during early ideation when outputs offered extractable, adaptable, or executable elements. Low feasibility, excessive cost or risk, generic content, and simultaneous factual and faithfulness violations inhibited inspiration. For \emph{suspected factual hallucinations}, whether truth was task-critical determined whether inspiration preceded verification or emerged from real materials found during it. Low-cost verification and partial truth content supported the latter path, whereas baseless claims could expose inspiring gaps only when truth was not task-critical. \emph{Confirmed factual hallucinations} rarely inspired unless creative elements could be separated from the falsehood. Faithfulness hallucinations inspired participants only when core constraints were preserved, while inspiration more likely emerged with departures from subjective constraints, especially in creative tasks with weak factual boundaries. Hypothetical recoding further showed that these effects depended on the user's goal, task stage, and the hallucinated content.

\begin{table*}[t]
  \centering
  \caption{Factors shaping whether participants perceived LLM hallucinations as inspiring. "Both" denotes factors shared across hallucination types; "Fact." and "Faith." denote factual and faithfulness hallucinations, respectively. A "+" indicates that a factor supported inspiration, whereas a "-" indicates that it inhibited inspiration. "Num." denotes number of participants.}
  \label{tab:inspiration-factors}
  \begin{tabular}{llcc}
    \toprule
    Type & Factors & Effect & Num. \\
    \midrule
    Both & Goal relevance (necessary) & + & All \\
     & Early-stage ideation & + & 19 \\
     & Departures can be transformed into creative elements by users & + & 13 \\
     & Low feasibility, excessive cost, or safety risks & - & 6 \\
     & Generic content & - & 5 \\
     & Simultaneous factual and faithfulness violations & - & 6 \\
    \midrule
    Suspected Fact. & (truth is task-critical) Useful discoveries during verification & + & 13 \\
     & (truth is task-critical) Low-cost verification & + & 9 \\
     & (truth is task-critical) Partial truth or traceability & + & 4 \\
     & (truth is not task-critical) Gaps exposed after verifying baseless claims & + & 7 \\
    Confirmed Fact. & Only when containing unexpected creative elements & + & 7 \\
    \midrule
    Faith. & Preserved \emph{core constraints} (necessary) & + & 13 \\
     & Departure from \emph{subjective constraints} & + & 13 \\
     & Creative tasks with weak factual boundaries & + & 9 \\
    \bottomrule
  \end{tabular}
\end{table*}

\subsubsection{Shared Factors Across Factual and Faithfulness Hallucinations}\label{sf:factors-share}

Goal relevance was necessary for all participants, while 19 (all except P16) said that an unrelated hallucination would not inspire them. This boundary applied to both hallucination types. P1 would keep exploring a fabricated paper if it was "very relevant to my research question in method, theory, or application." P15 likewise judged inspiration by whether the content was "related to my goal." Conversely, P2 ignored extra content with "little to do with my current task." Hypothetical recoding reinforced this pattern. Twelve participants changed green cases to yellow or red when another goal made the same content meaningless. Nine changed yellow or red cases to green when a more relevant goal made the departure actionable. P8 noted LLM-fabricated \emph{Butterfly Lovers} storyline suited a personal cultural product but not an urban product that had to reflect the real story. P15, for instance, mentioned LLM-generated magnetic charging design violated the existing circuit layout but would become a good idea if the budget allowed a circuit redesign.

Stage of work also shaped inspiration. Nineteen participants (all except P19) were more willing to accept hallucinations during ideation than execution. Across research, design, and content creation, early thinking, topic selection, and brainstorming made departures from facts or constraints more useful. P2 accepted "bold ideas that have not necessarily been fully verified" during early design or concept generation. P6 said factual hallucinations during topic exploration might "trigger me to search for information and discover a direction that nobody has tried." P17 reported the same pattern in script writing:
\begin{quote}
    "faithfulness hallucinations in early plot setting and character relationships could help me move beyond the limits of my original idea." 
\end{quote}
During execution, participants more often treated hallucinations as risks or errors. They were particularly unwilling to accept them in research writing and translation (P4/7/11) or prototyping and engineering design (P10/15/20), and finalizing, or reviewing content (P6/9/14/18).

Thirteen participants considered departures that can be transformed into creative elements are central to inspiration. Participants extracted classification schemes from fabricated data or papers (P2/7) and retained the "embroidery experience" module from an incorrect travel recommendation (P14). Others adapted a curved surface from a noncompliant shell design (P15) or repaired illogical plots while retaining useful actions or story arrangements (P17/18). Conversely, six participants (P3/6/9/13/15/20) said that low feasibility, excessive cost, or risk suppressed inspiration even when a hallucination seemed interesting. P6 requested visually striking food-video ideas that were safe and feasible at home, but the LLM suggested dry ice. Because this introduced safety risks and offered no transformable visual-impact strategy, P6 coded the faithfulness hallucination red and said that no alternative scenario would change its color.

Five participants (P1/11/13/16/17) found generic content uninspiring because it did not provide the transcendence required for inspiration. Examples included familiar paper-writing advice (P1/11) and plot routines (P16/17). Hypothetical recoding showed that this content remained uninspiring across contexts. When the LLM ignored P11's recording and instead advised "standardizing the writing or figures", P11 considered the faithfulness hallucination "too common" to inspire in any situation.

Finally, six participants (P1/3/5/8/14/18) reported outputs with both factual and faithfulness violations unlikely to inspire because they lacked an factual basis and violated the user's requirements. Five had encountered similar cases, and P18 provided one from a murder-mystery script. The LLM wrote that a character feared an inspector would discover his illegal resale after the inspector had died. P18 coded the output red because, without a supernatural setting and under a logical-reasoning requirement, the dead inspector could not investigate and the plot violated the requested logic. P18 said that no alternative condition would make the case yellow or green.

\subsubsection{Factors Specific to Factual Hallucinations}\label{sf:factors-fact}

Together with participants, we distinguished \emph{suspected factual hallucinations} from \emph{confirmed factual hallucinations}. The former were claims whose truth participants did not know without verification; the latter were claims they could immediately recognize as false.

For \emph{suspected factual hallucinations}, the first judgment was whether truth affected task progress. When the factual truth was not critical to the task (truth was not task-critical), seven participants (P2/6/8/10/14/15/20) could be inspired directly. P2, for example, suspected that the generated numerical data were fabricated which they did not plan to use, but the data classification scheme suggested P2 how to measure the data. For P6, LLM-fabricated viewing data did not determine whether "rice-paper tofu pizza" could be filmed, while the dish itself was novel and attention-grabbing. When the task depended on whether something existed (truth was task-critical), 10 participants verified first (P1/2/4/7/10/12/14/17/19/20). This included papers for academic writing (P4/7) and art-historical information for P19's course video.

Therefore, \emph{Suspected factual hallucinations} followed two inspiration routes. When truth was not task-critical, inspiration preceded verification in Route 1. When truth was task-critical, inspiration arose from other real and useful materials found during verification in Route 2. Content verified as true fell outside our scope; every \emph{suspected factual hallucination} discussed here was ultimately confirmed by participants as a factual hallucination.

Here, we firstly focus on route 2. Thirteen participants treated \emph{suspected factual hallucinations} as indirect sources of inspiration. When suspicious content was task-relevant, participants verified it, and the search could surface real papers, artworks, locations, or other examples even if the original item was fabricated. P4 said that searching for fabricated but related papers could "open up my thinking and make me explore further." After the LLM falsely claimed that medieval Europeans used spice clocks to tell time by smell, P12 found East Asian incense clocks that measured time by the degree of incense burned when verifying spice clocks, inspiring a video topic.

This also means that verifiability and verification cost shaped whether \emph{suspected factual hallucinations} inspired participants in Route 2 (P2/4/6/8/10/12/14/19/20). Participants preferred suspected contents that could be checked quickly via search engines, map services, Google Scholar, social media, or LLM-provided links, especially when verification could yield goal-relevant material. The hallucination was fabricated without any factual basis (we call it baseless hallucination), by contrast, were almost meaningless. While developing an art-education topic, P19 checked the LLM's claim \texttt{"The sculptor [...] carved a Poseidon statue in Copenhagen in 1915."} P19 found neither the sculptor nor such a Copenhagen statue, thus coded the case red, and said that it could never become green or yellow. Therefore, partly true or traceable hallucinations were more likely to support inspiration. Examples included fabricated data with useful categories (P2), papers with real research problems (P4), fake videos pointing to real design issues (P10), and incorrect stores suggesting real service experiences (P14).

In contrast, in route 1, even a baseless hallucination could inspire when truth was not task-critical. Seven participants (P1/4/6/10/12/14/20) viewed this gap as a possible opportunity. For P1 in research, fabricated studies matching a research idea could reveal that the research framing needed reconsideration since the absence of real studies. For P20 in game design, a fabricated mod and immature gameplay idea suggested a gap in the mod ecosystem. For P6 in content creation, although the viewing data for a creative food idea was verified to be fake, the absence of similar food videos also made the topic appear promising.

For \emph{confirmed factual hallucinations}, ten participants (P2/3/4/5/7/9/13/17/18/19) found these hallucinations usually uninspiring. Because the directly recognized falsehood offered nothing cognitively new and therefore lacked transcendence. However, seven participants (P6/8/10/12/14/15/20) could still be inspired only when a confirmed falsehood contained an unexpected, transformable creative element (Route 3). P8, for example, recognized that the LLM's altered \emph{Butterfly Lovers} story was false, but its fabricated character relationships better suited an umbrella design and were eventually adopted.

\subsubsection{Factors Specific to Faithfulness Hallucinations}\label{sf:factors-faith}

For faithfulness hallucinations, inspiration depended on which constraint was violated. Thirteen participants distinguished \emph{core constraints}, which determined whether a task remained viable, from \emph{subjective constraints}, which reflected adjustable preferences or requirements. Core constraints included safe visual impact (P6), avoidance of stimulating or frightening content (P9/12), and specified materials (P11), keeping the flower as the main object (P13), an existing circuit layout (P15), and mystery deduction unaffected by magic (P18). P6 said that violating \emph{core constraints} "generally cannot inspire me," while P18 said that crossing such a bottom line "has no meaning to me." Preserving \emph{core constraints} was therefore necessary for a faithfulness hallucination to inspire.

Thirteen participants said that departures from non-essential or \emph{subjective constraints} could inspire when they remained task-relevant. P9's request excluded familiar sleep-aid elements such as white noise, yet the faithfulness hallucination around "delivering dreams" inspired a new direction. P13 requested a rendering style, but the image model produced a grass-ring-like form that better fits their visual concept. The color-coding results further showed that green cases usually contained transformable and executable elements (P3/5/9/11/13/15--18), such as a better curved grip design that departed from P15's example image, or plot and alibi elements that broke logic but enriched the mystery script for P17 and P18. Attractive or playful departures that did not support the task tended to be yellow (P9/13/16). P9, for example, found "clouds sleeping in a glass bottle" suitable for a healing video but not the prompted abstract sleep-aid video. These results echoed the shared factors in Section \ref{sf:factors-share}.

Faithfulness hallucinations were especially salient in fictional or generative tasks such as visual design and short-video scripting, where factual boundaries were weaker and participants cared more about whether the output followed their settings and requirements (P3/5/8/9/13/15--18). Since outputs often exceeded participants' prior imagination, inspiration depended less on factual correctness than on whether a departure revealed a better possibility. P15 explained that users "cannot fully know whether their existing requirements are reasonable or whether constraints block better outputs." When P3 asked the LLM to follow a railway-equipment image, it added wheels and connection modules absent from this reference. Although this violated the prompt, the images "made me realize that the LLM could generate equipment details for better demonstration effects" (P3).

\subsection{User Behaviors}\label{sf:behaviors}

This subsection addresses RQ2 by describing what hallucinations inspired participants to do and how they responded when no inspiration occurred. For factual hallucinations, whether truth was task-critical determined three routes. Participants explored then verified (Route 1), verified and drew inspiration from real evidence (Route 2), or retained creative elements from known falsehoods (Route 3). Participants otherwise ignored or corrected the output, requested another answer, or supplied materials. Inspiring faithfulness hallucinations led participants to use or develop content, reorganize constraints, or repair faithfulness problems. Non-inspiring cases prompted them to revise the prompt, generate another response, or restart a conversation (Figure \ref{fig:behavior-framework}).

\subsubsection{Behaviors After Factual Hallucinations}\label{sf:behaviors-fact}

As described above, factual hallucinations led to three behavior routes. A \emph{suspected factual hallucination} inspired participants before verification in Route 1 or through verification in Route 2. In Route 3, a \emph{confirmed factual hallucination} inspired only through separable creative elements.

In Route 1, 12 participants could be inspired by useful, unfamiliar content before verification when truth did not determine immediate task success. They recorded or explored the idea, sometimes continuing the LLM conversation, and verified it afterward. P2 "wrote it down first" and planned to measure the data using the suggested categories. Before opening fake video links, P10 found the summaries of product-design gaps useful for a report, recorded them, and later checked which gaps others had actually discussed.

In Route 2, goal-relevance motivated 11 participants to verify \emph{suspected factual hallucinations} before inspiration because truth was task-critical. Related real materials found through searches could then inspire them. P4's inspiration "came more from the real papers I found after verifying LLM-fabricated paper." P19 distrusted generated art history but often found inspiration in real artworks during verification. Factual hallucinations thus served as fuzzy search entry points. Participants then incorporated the real material into their work or continued discussing it with the LLM.

In Route 3, participants usually ignored \emph{confirmed factual hallucinations} unless they contained unexpected creative elements that could be used in their work and thereby bring inspiration. (see Section \ref{sf:factors-fact}). Seven recorded these creative elements or continued discussing them with the LLM. As P6 explained, "If something is obviously false but contains a useful combination, I may throw away the factual part and keep only the creative structure."

When factual hallucinations did not inspire, 17 participants ignored them, requested another answer, corrected the LLM, or supplied factual materials. After verifying a generated paper as fake and finding nothing new, P1 said, "I usually tell the LLM that I did not find the related study." P1 ended the chat if later chats remained useless. P20 likewise explicitly told the LLM that a generated game mod did not exist.

\subsubsection{Behaviors After Faithfulness Hallucinations}\label{sf:behaviors-faith}

Inspiring faithfulness hallucinations prompted three behaviors in the \emph{inspired-to} state. First, ten participants directly used or developed the content, such as using a constraint-violating generated image in demonstration (P3) and adapting the generated curved surface which violate provided example image for a future product (P15). Second, nine reorganized prompt constraints. P14 considered relaxing constraints when all requirements could not be met, whereas P5 asked why the LLM had violated a constraint. Third, four participants in script-writing and visual-design tasks (P9/16--18) repaired only the parts with faithfulness problems while preserving the overall content. P9 converted "delivering dreams" into abstract light spots for a sleep-aid video, and P17 added low blood sugar to explain why a victim drank poisoned sugar water. P18 summarized this pattern:
\begin{quote}
    "Many times the LLM gives me a good script base even though some parts are illogical.Such scripts will encourage me to keep refining them so that they become more suitable for role‑playing social games."
\end{quote}

Regardless of inspiration, all participants said that faithfulness hallucinations prompted them to reconsider the prompt: whether requirements were clear, materials were missing, or the LLM had misunderstood the task. After non-inspiring cases, 12 explicitly checked and revised the prompt before requesting another answer. P13 and P20 might delete the chat and start again if the hallucination had contaminated the conversational context.

\subsection{Cognition and Experience}\label{sf:cognition}

This subsection addresses RQ3 by examining how hallucination-inspired experiences shaped participants' views of hallucination, perceived inspiration under changing constraints, and trust in different LLM roles.

\subsubsection{Hallucinations Versus Non-Hallucinatory Suggestions}\label{sf:normal-suggestions}

Almost all participants agreed that non-hallucinatory LLM suggestions could more readily lead to direct inspiration. By contrast, 14 participants described hallucination-inspired experiences as more unexpected and said they opened previously unimagined directions, especially during early ideation and brainstorming. P15 captured this contrast: "Whether hallucinations can bring inspiration is more extreme, being either eye-opening or terrible. Normal suggestions are more stable and controllable, like an expected surprise." Compared with non-hallucinatory suggestions, inspiring hallucinations could reveal an unexpected angle, category, story mechanism, visual image, or unmet need. Eight participants also reported additional effort and interaction costs: factual hallucinations required verification, whereas faithfulness hallucinations required users to extract and reshape the creative idea. Only three participants found little difference between the two experiences.

\subsubsection{Perceived Inspiration Under Changing Constraints}\label{sf:changing-constraints}

As noted in Section \ref{sf:factors-faith}, participants distinguished \emph{core constraints} from \emph{subjective constraints}. Consistent with prior work on inspiration seeking through LLMs \cite{lin_seeking_2025}, users often began conversations without knowing all their constraints. Then participants gradually adjusted these constraints in response to generated content and faithfulness hallucinations. As constraints accumulated, faithfulness hallucinations generally became less likely to inspire.

In this process, six participants (P4/7/9/10/13/20) said that faithfulness hallucinations were more likely to inspire early in a conversation, when constraints were fewer or less specific. At that stage, the LLM might not yet understand their intentions, while loose constraints could also enable more imaginative ideas. P4 said that during early ideation, a task-related idea could remain useful even if it violated the prompt when constraints were relaxed. One participant explained:
\begin{quote}
    "At the beginning of making a video, I often had not yet decided what the next video should be, so there were fewer important requirements. A faithfulness hallucination that violated personal preferences was therefore more likely to inspire me" (P9).
\end{quote}

As constraints increased, 12 participants believed that faithfulness hallucinations became more likely to occur. Eight attributed this pattern to requirements accumulating over a conversation: excessive constraints could distort user intent, make compliance harder, and increase the chance that the LLM omitted requirements. For instance, complex research questions and requirements made complete answers harder (P1) and no real travel route might satisfy every condition (P14). P14 also said that factual hallucinations could increase when the LLM fabricated nonexistent information to satisfy all conditions.

Meanwhile, 15 participants said that faithfulness hallucinations usually became less likely to inspire as constraints accumulated. Participants added stricter constraints as their goals and requirements became clearer. Hallucinations then became more likely to violate \emph{core constraints} or the task goal, reducing usability. After developing a sleep-aid video idea, P9 added many "do not" requirements, but the LLM kept generating those elements, making its suggestions meaningless. Therefore, P13 explained: "For faithfulness hallucinations, fewer constraints make inspiration more likely, and more constraints make inspiration less likely, because as I add constraints, I am also gradually fixing my design plan. At that point, I just want the model to follow the requirements." P15 similarly said that violations almost never inspired once constraints and needs became strict, especially when engineering details could no longer change.

\subsubsection{Trust and Role Differentiation}\label{sf:trust}

Participants reported that hallucination-inspired experiences changed their trust in different LLM roles. Five (P5/7/8/10/14) said that factual hallucinations reduced the perceived reliability of the LLM or a specific answer. Four (P2/4/12/19) did not trust LLM-generated content because LLMs often produced false information. Eight (P6/8--10/13/17/18/20) said that faithfulness hallucinations reduced trust in the LLM's constraint adherence. Thus, both types could reduce trust even when they later became inspiring.

Yet inspiring hallucinations could also strengthen trust in other LLM roles. Eleven participants said that factual hallucinations maintained or reinforced verification habits, especially when truth was task-critical. Separately, after factual hallucinations inspired them, 11 participants positioned LLMs more strongly as divergence or fuzzy search tools because verification could reveal additional novel content. P6 said, "I would trust the LLM more as a creative divergence tool. But that does not make me trust its factual judgment more." similarly, P12 called the LLM "suitable as a topic partner, but not as a source of information." Faithfulness hallucinations also helped clarify constraints and increased some participants' willingness to use LLMs as early-stage co-ideation partners. Eight participants (P3/5/6/9/13/16/18/20) described an inspiring faithfulness hallucination as resembling a partner that deliberately departed from their request to offer something better. P9 considered the LLM suitable for early divergence after such inspiration. For P16, slight departures helped break fixed thinking in early, emotionally driven script writing with many cliched plots.

\subsection{Expectations for LLM Design}\label{sf:expectation}

This subsection addresses RQ4 by reporting what participants expected from LLMs when hallucinations arose and the design ideas they proposed to meet these expectations. All participants rejected deliberate hallucination generation, but not all wanted hallucinations eliminated completely. P10/12/20 did not want either factual or faithfulness hallucinations eliminated; P4/6/8 said this only of factual hallucinations; and P3/5/7/9/11/13/15 only of faithfulness hallucinations. These expectations reflected the preceding findings. P4 warned that if an LLM were proven to have no factual hallucinations, "users might lose the habit of verification, which would affect critical thinking." P15 considered faithfulness hallucinations inevitable because users cannot fully articulate every requirement and may interpret multi-constraint prompts differently from LLMs. Such hallucinations could also help users clarify their needs. Only P1/14/18 wanted hallucinations to disappear, while acknowledging that they were difficult to avoid.

\subsubsection{User Control}

In creative work, \textbf{almost all participants (all except P19) expected control over an LLM's divergence and adherence to evidence or requirements}. They proposed a creative or inspiration mode for choosing between bolder ideation and grounded responses. P2 wanted control over "the proportion between real or accurate content and bold ideas." P4 proposed "a parameter that makes the model more biased toward creativity or toward evidence-based replies." P7 wanted the LLM to "appropriately cross some constraints and give more directions" when divergence was desired. P20 suggested first stating what the LLM knew to be certain and then separately presenting what it would imagine when users wanted the LLM to diverge.

\textbf{Six participants (P3/6/8/9/12/15) expected LLMs to sense the stage of creative work.} During early brainstorming, they wanted imaginative ideas that could depart from facts or parts of the prompt. After clarifying their goals and methods and beginning execution, they did not want hallucinations. Their corresponding design idea was stage-sensitive control that enabled early divergence without making hallucination the default.

For faithfulness hallucinations, 5 participants (P7/9/14/18/20) expected direct, manual control over constraint priorities. They proposed marking \emph{core constraints} that must not be violated and \emph{subjective constraints} that could be handled flexibly, because violations of core constraints rarely inspired them. P9 wanted prompt annotations specifying "which constraints cannot be violated and which ones I think can be handled flexibly." P18 wanted to "lock core settings." This design expectation would make constraint priorities explicit rather than leaving the LLM to infer them.

\subsubsection{Hallucination Identification, Labeling, and Mitigation}

Participants expected identification, labeling, and mitigation to reflect the different references violated by factual and faithfulness hallucinations. Their ideas combined shared and type-specific mechanisms. Across both types, they generally wanted identified content highlighted or labeled without full outputs being changed, so that the text or artifact remained readable and users could decide what to do next.

For factual hallucinations, 13 participants expected support for reducing, identifying, or labeling them. One widely used mechanism, available in LLMs such as ChatGPT\footnote{\url{https://chatgpt.com/}}, provides links or sources for rapid verification. Beyond links, \textbf{10 participants wanted LLMs to label content lacking sources or links}. P12 wanted a distinction between "factual basis" and "creative association." P6 wanted labels for "what is fact, what is inference, what is a creative hypothesis, and what needs verification." For example, divergent ideas could be marked "speculation" or "creative idea," while potentially fabricated data tables could receive low-confidence labels. \textbf{Nine participants also expected LLMs or external tools to check factuality automatically}, by checking whether linked sources supported claims and assigning confidence ratings (P20).

For faithfulness hallucinations, participants noted that an LLM might not know whether its output violated user constraints. \textbf{Seven (P3/5/6/9/11/17/18) expected LLM self-checks or external identification techniques.} They suggested another AI system or hallucination benchmark (P3/5/6/9/11), or explicit LLM self-evaluation: "Asking the LLM to evaluate whether a generated plot met my requirements, whether the logic was coherent, and what needed correction" (P17/18). \textbf{P13 expected LLMs to learn how individual users judge whether generated content violates their requirements}, because faithfulness hallucinations do not always have an objective evaluation standard. \textbf{Six participants (P1/7/14/15/18/20) expected the LLM to disclose when constraints could not all be satisfied rather than force an answer.} For idealized goals or complex constraints, it should identify requirements that might need relaxation and offer options instead of fabricating facts or quietly changing requirements to produce a pleasing response.

\subsubsection{Salvage Support}

Reflecting on how they had turned useful elements from unavoidable but potentially inspiring hallucinations into practice, some participants expected LLMs to support this process. We call this design expectation \emph{salvage support}: preserving, repairing, and developing useful parts of a hallucination without carrying forward harmful ones.

\textbf{Three participants (P7/13/17) expected local editing, annotation, pinning useful content, and preservation of work traces.} They imagined LLM interfaces where users could check, delete, cross out, or annotate output segments. The system would use these edits as prompts in later conversations to correct hallucinations and remember creative ideas. P7 wanted to "mark the wrong content and pin the useful content." P17 said, "I hope I can first read it myself, tell the model which parts I think are problematic, and then let the model find similar hallucinations or logical problems."

Five participants (P6/9/14/17/18) expected a more ambitious form of automatic creative element extraction support. After a hallucination was identified or marked by the LLM or user, they did not want it merely labeled or deleted. \textbf{They wanted the LLM to explain which parts of the hallucination were unusable and which could be transformed into safe, truthful, or constraint-satisfying plans.} P6 wanted LLM to "preserve useful uncertainty while blocking danger and misinformation." P9 wanted it to identify constraint-violating but potentially creative content and "help me extract the transformable part." Thus, \emph{salvage support} aims to support inspiration while correcting errors.

\section{Discussion}

This Discussion first presents a framework that connects the factors shaping whether factual and faithfulness hallucinations inspire creative practitioners with users' subsequent behaviors. We then discuss how inspiration from faithfulness hallucinations changes across stages of creative work and how hallucination experiences reshape trust in specific LLM roles. Next, we argue that evaluations of hallucinations and LLM creativity should account for users' subsequent judgments and actions. We also summarize LLM designs with stage-sensitive controls, explicit constraint priorities, hallucination labeling and mitigation, and salvage support for the \emph{inspired-by} and \emph{inspired-to} states. Finally, we conclude the study's limitations and future work.

\subsection{From Hallucination to Inspiration}\label{sd:hallu4ins}

\begin{figure}
    \centering
    \includegraphics[width=1\linewidth]{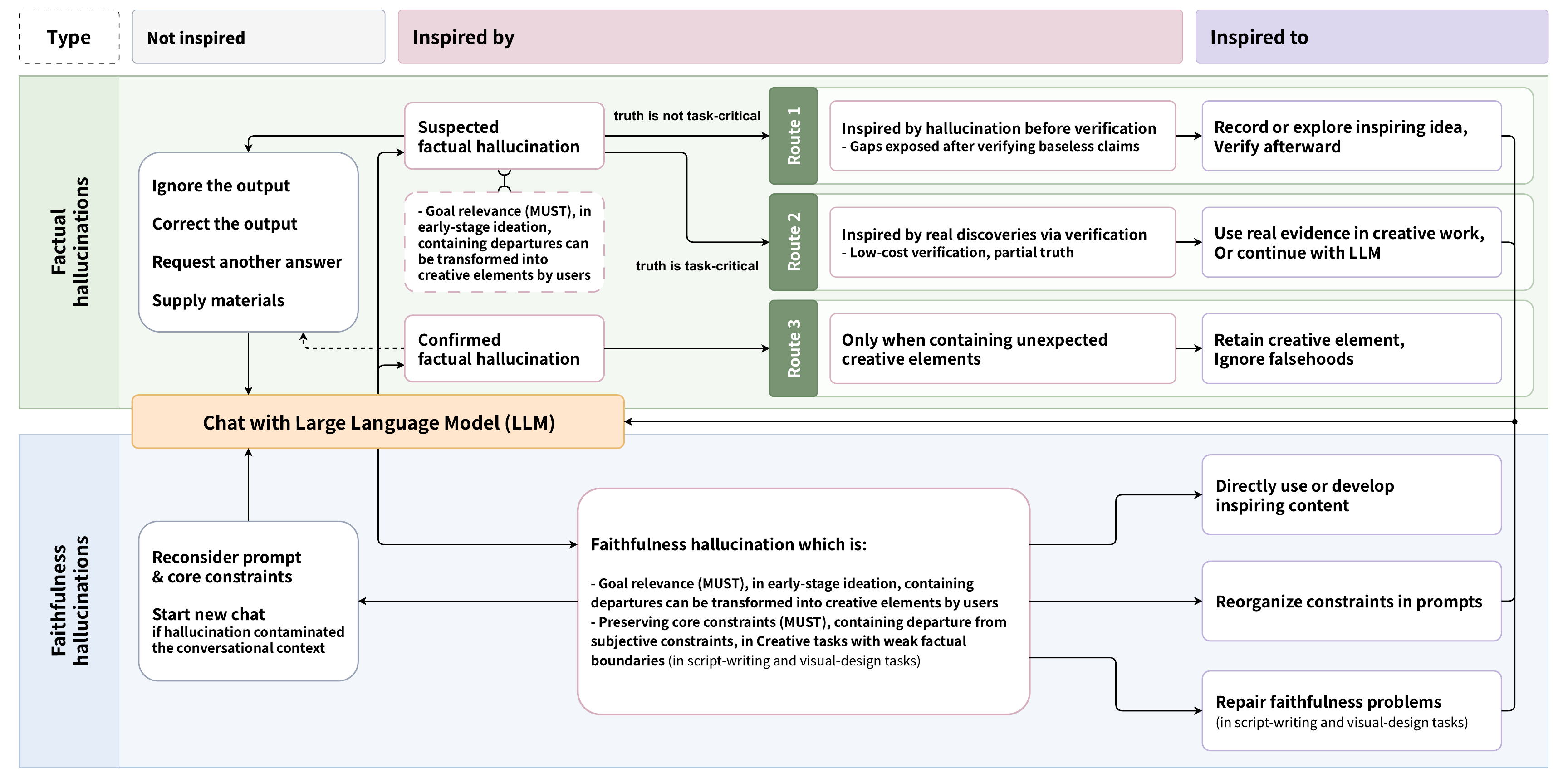}
    \caption{Framework from hallucination to inspiration in creative work}
    \label{fig:behavior-framework}
    \Description{A flowchart has two horizontal bands for factual and faithfulness hallucinations and four columns labeled Type, Not be inspired, Inspired by, and Inspired to. In the factual band, suspected factual hallucinations follow Route 1 when truth is not task-critical and Route 2 when truth is task-critical. Route 1 moves from inspiration before verification to recording or exploring an idea and verifying it afterward. Route 2 moves from verification to inspiration from real discoveries and then to using that evidence or continuing with the LLM. Confirmed factual hallucinations follow Route 3 only when they contain unexpected creative elements, which users retain while ignoring the falsehood. Non-inspiring factual hallucinations lead users to ignore or correct the output, request another answer, or supply materials. In the faithfulness band, hallucinations inspire when they remain goal-relevant, occur during early-stage ideation, preserve core constraints, and contain transformable departures, especially departures from subjective constraints in creative tasks with weak factual boundaries. These hallucinations lead users to develop the content, reorganize prompt constraints, or repair faithfulness problems. Non-inspiring faithfulness hallucinations lead users to reconsider the prompt or start a new chat. Feedback arrows connect these behaviors to continued interaction with the LLM.}
\end{figure}


Thrash et al. characterize inspiration through evocation, transcendence, and approach motivation, distinguishing appreciation of a possibility in the \emph{inspired-by} state from action to realize it in the \emph{inspired-to} state \cite{thrash_inspiration_2003,thrash_inspiration_2004}. Lin et al. extend this account to human-LLM interaction through a proactive, iterative cycle in which users refine incomplete preferences, become inspired by personalized and unexpected suggestions, and then search, verify, extend, or evaluate; discoveries during this process may provide further inspiration \cite{lin_seeking_2025}. In addition, research on hallucinations has examined how hallucinations reduce output quality and alter trust or subsequent interaction, as well as how users respond to their negative effects by recognizing, comparing, verifying, and correcting hallucinated content \cite{nahar_fakes_2024, cheng_relic_2024, leiser_hill_2024, ashktorab_emerging_2025, lin_seeking_2025, ryser_calibrated_2025}. In creative work, studies suggest that although hallucinations can disrupt creation, these unexpected errors may also provide new ideas and prompt reflection \cite{jiang_survey_2024, he_shakespearean_2025, holmstrom_what_2026}. Extending this view, Lan et al. use conceptual analysis to argue that hallucinations may support creativity when they bring context-relevant novelty and people can use it \cite{lan_schizosyntax_2026}. Taken together, these studies show that hallucinations may open new creative directions, inspire users, and affect creative work, but they do not explain how creative practitioners judge hallucinations to be inspiring in the \emph{inspired-by} state, or how that content and judgment motivate \emph{inspired-to} behavior.

Our findings address this gap through the framework in Figure \ref{fig:behavior-framework}. Drawing on Sections \ref{sf:factors} and \ref{sf:behaviors}, the framework connects the factors shaping inspiration with subsequent behaviors across two hallucination types and three states: not-inspired, \emph{inspired-by}, and \emph{inspired-to}. \textbf{At its center is users' situated interpretation of hallucinations: when an LLM generates hallucinations, practitioners judge whether it is inspiring in relation to their task stage, knowledge, and circumstances, provided that it exceeds their prior knowledge and remains relevant to their goals.} This judgment branches by hallucination type. For factual hallucinations, users' understanding of the content's truth status and whether truth is task-critical determine where inspiration lies. A \emph{suspected factual hallucination} may inspire through a possibility explored before verification or through real discoveries made during verification, whereas a \emph{confirmed factual hallucination} can inspire only through the unexpected creative elements it contains. Then, Route 1, Route 2, and Route 3 lead to different \emph{inspired-to} behaviors. Faithfulness hallucinations follow one main path: when \emph{core constraints} remain intact, users may be inspired by departures from \emph{subjective constraints} in relation to their own needs and those of the task, which can lead to three \emph{inspired-to} behaviors. For both hallucination types, users continue their dialogue with the LLM after a non-inspiring response or after entering the \emph{inspired-to} state with a new possibility, until they finish the creative work. Overall, the framework shows the relationship and cycle between situated judgment and behavior in the process from hallucination to inspiration.

Moreover, our framework contributes three points to HCI's understanding of hallucinations and inspiration. First, Lin et al.'s framework begins with proactive inspiration seeking and explains how users refine their preferences to obtain personalized suggestions that inspire them \cite{lin_seeking_2025}. In comparison, our framework locates inspiration primarily in users' judgments of hallucinations they did not actively seek, focusing on factual status for factual hallucinations and constraint priorities for faithfulness hallucinations. Second, the framework provides a basis for analyzing creative processes in studies and evaluations of co-creative LLM systems after an LLM generates a hallucination \cite{karimi_evaluating_2018}. For studies, it traces the full trajectory from hallucination through situated judgment to inspiration and subsequent action, helping researchers examine which inspiration state users reach and what work connects \emph{inspired-by} with \emph{inspired-to}. For evaluations, it shows how users find inspiring content within or through hallucinations, what they do next, and the potential interaction costs of moving from hallucination to inspiration. Together with the hypothetical recoding results, the framework indicates that hallucinations can have creative value depending on users' situated interpretations; they therefore should not be treated as uniformly harmful without considering users' judgments in different circumstances, as discussed in Section \ref{sd:evalhallu}. Third, the framework provides theoretical support for interaction designs that help LLMs respond to hallucinations in ways that preserve their inspirational potential while avoiding harm, as discussed in Section \ref{sd:design}.

\subsection{Situated Inspiration and Trust}

\subsubsection{Complementary Paths to Inspiration}\label{sd:compath}

Seeking inspiration through LLMs usually involves actively clarifying one's preferences: as requirements become more specific, non-hallucinatory suggestions are more likely to exceed users' prior knowledge while fitting their needs \cite{lin_seeking_2025}. Section \ref{sf:changing-constraints} shows a different pattern for faithfulness hallucinations, which were more likely to inspire early, before requirements were fully formed, but participants later perceived them as more likely to occur and less likely to inspire. This pattern does not imply that fewer early constraints are always better. In image inspiration searches, different constraint levels elicit different search strategies, and material found under low constraints is not necessarily the most inspiring \cite{biskjaer_how_2020}. The two paths are therefore better understood as complementary across creative stages. Early non-hallucinatory suggestions may be too generic for an unclear need, whereas a modest, goal-relevant departure from user requirements can sometimes open a direction. Later, non-hallucinatory suggestions can draw on established preferences, while faithfulness hallucinations are more likely to disrupt requirements that users have settled. This staged relationship motivates our recommendation that LLMs adjust divergence and adherence over a task (in Section \ref{sd:design}).

\subsubsection{Role-differentiated Trust}

Hallucination research often associates errors with reduced trust, over-reliance, and poor decisions \cite{huang_survey_2025,leiser_hill_2024, ye_cognitive_2023, li_lokis_2025, zhang_sirens_2025}. Our participants likewise became less willing to trust an LLM's factual judgments and constraint adherence (Section \ref{sf:trust}), but this did not necessarily amount to an overall loss of trust. After an inspiring factual hallucination, they were more willing to use the LLM for early search while still verifying its claims; when a faithfulness hallucination opened a new direction, they were more willing to use it as an ideation partner, but not to rely on it during execution. Prior work shows that users calibrate trust in hallucinated output according to risk, context, experience, and domain knowledge \cite{ryser_calibrated_2025}; our findings suggest that this calibration shapes not only how much users trust an LLM, but also which tasks they assign to it. HCI research should therefore distinguish trust in an LLM as a source of facts, a follower of requirements, a source of unexpected directions, or an ideation partner. Overall, hallucinations did not simply raise or lower trust; they led users to redraw which LLM roles they trusted.

\subsection{Evaluation of Hallucination and Creativity}\label{sd:evalhallu}

Most current hallucination evaluations use benchmarks \cite{li_halueval_2023, bang_hallulens_2025, ren_detecting_2026} or techniques such as semantic entropy \cite{farquhar_detecting_2024} to determine whether an output departs from facts, source material, input context, or a reference answer. Although indispensable for identification, these approaches evaluate the generated content itself. Our findings and prior research indicate that a hallucination remains an error, but its consequences vary across users and tasks and are not uniformly perceived as harmful. Hallucinations can produce low-quality data \cite{ashktorab_emerging_2025}, confuse learners \cite{steinbach_when_2025}, or cause clinical harm \cite{asgari_framework_2025}, yet they may also support divergent thinking during learning \cite{holmstrom_what_2026} or serve as a potential source of inspiration. Participants' hypothetical recoding (Section \ref{sf:factors}) illustrates this context dependence: the same hallucination may open a direction during early ideation but create risk during execution or publication, and may be useful for one goal but meaningless for another.

These findings suggest that hallucination evaluation should distinguish error identification from assessment of its consequences. To our knowledge, current evaluations have not treated downstream interaction consequences as a separate dimension. \textbf{We therefore propose supplementing evaluations of whether an output is a hallucination with an assessment of what follows in interaction.} In creative tasks, this assessment could record the user's goal and task stage; distinguish violations of external facts, \emph{core constraints}, and \emph{subjective constraints}; and examine whether the hallucinated content bring new ideas or increased the risk of accepting false information and the burden of correction. This dimension neither reclassifies errors as inspiration nor relaxes factuality and faithfulness standards; it avoids inferring interaction quality from hallucination rates alone.

This proposal also applies to creativity evaluation in co-creation systems. Co-creative LLMs should be evaluated not only through final artifacts but also through the creative process and human-LLM interaction, with attention to who evaluates and when \cite{karimi_evaluating_2018}. As Section \ref{sd:hallu4ins} argues, the path from hallucination to inspiration can be understood as a co-creative process: value is not determined by the output alone, but emerges as users judge it. Many evaluations of LLM creativity instead compare model and human outputs through divergent-thinking tasks and measures of novelty, appropriateness, or diversity \cite{bellemare-pepin_divergent_2026, he_shakespearean_2025, banerjee_does_2026}. These measures indicate whether an output is itself creative, but not how LLM creativity supports user creativity, and reveal little about hallucinations in the user's creative process. Including interaction consequences could broaden co-creative LLM evaluation and help HCI researchers examine how hallucinations shape creative work rather than only generated outputs.


\subsection{LLM Designs for Hallucinations and Inspiration}\label{sd:design}

Prior inspiration-oriented design recommendations emphasize helping users express and update preferences and understand why generated content relates to their goals \cite{lin_seeking_2025}. Building on participants' design expectations in Section \ref{sf:expectation}, the framework in Section \ref{sd:hallu4ins}, and discussion in Section \ref{sd:compath}, our findings suggest: in the \emph{inspired-by} state, LLMs should provide stage-sensitive control over divergence and adherence, user-defined constraint priorities, differentiate factual from faithfulness hallucinations in identification, labeling, and mitigation, and salvage creative elements from detected hallucinations; in the \emph{inspired-to} state, they should provide user-directed \emph{salvage support}.

\subsubsection{Support the Inspired-by State}

\textbf{First, LLMs should let users adjust output divergence and adherence to their requirements.} Existing co-creative tools support exploration along user-defined dimensions or control over story generation \cite{almeda_prompting_2024, chung_talebrush_2022}, while experiments show that the effects of divergent and convergent LLM roles vary across users and stages \cite{rosenbaum_scaffolding_2025}. Together, these studies provide precedents for exposing such control, but do not explain how an LLM should adapt the balance over a creative task. Combining participants’ expectations regarding user control and the explanation of stage-sensitive adjustment in Section \ref{sd:compath}, we suggest providing a mode: an LLM could increase divergence - either automatically or under user control - during early ideation and brainstorming while labeling speculation or constraint violations, then prioritize adherence after the user has settled a goal and requirements and begun realizing the idea. This recommendation aims to align output strategy with the changing needs of creative work.

\textbf{Second, LLMs should make constraint priorities explicit and user-defined.} ConstraintMaker lets users select, edit, and combine output-format constraints through a graphical interface, while separating the task description from the expected format so constraints remain visible and adjustable \cite{liu_we_2024}. Our participants further proposed marking requirements as \emph{core constraints} that must not be violated or \emph{subjective constraints} that may be handled flexibly. An LLM should always respect core constraints, but could depart from subjective constraints when this might reveal a better possibility. This design leaves constraint importance to users, reducing faithfulness hallucinations that undermine the task while preserving room for inspiration from departures they explicitly allow.

\textbf{Third, our findings extend design implications for hallucination identification, labeling, and mitigation.} Prior work shows that warnings can improve users' recognition of hallucinations \cite{nahar_fakes_2024}, while RELIC and HILL compare multiple generations and highlight suspicious content \cite{cheng_relic_2024,leiser_hill_2024}. Beyond classifying or highlighting hallucinations and linking sources, participants' proposals (Section \ref{sf:expectation}) distinguish factual from faithfulness hallucinations. For factual hallucinations, interfaces could separately label source-supported information, low-confidence claims, speculation, and creative hypotheses, helping users decide what needs verification and whether it is LLM imagination. External tools could check whether claims, including claims with links, are supported and provide confidence estimates. For faithfulness hallucinations, an LLM could check output against explicit user requirements; revisions after this check may themselves inspire users (P17/18). It could also use external criteria and existing detection methods. Because faithfulness judgments can be subjective, an LLM should learn how each user decides that an output has departed from their requirements. When requirements cannot all be satisfied, it should disclose the conflict and ask the user to decide rather than silently ignoring one.

Finally, \emph{salvage support} (Section \ref{sf:expectation}) includes two suggestions for different inspiration stages. When the LLM has not detected a hallucination but the user has already been inspired, support belongs in \emph{inspired-to} and is discussed below. When the LLM detects a hallucination before the user judges it inspiring, support belongs in \emph{inspired-by}: rather than deleting the hallucinated content, participants proposed separating unusable parts from creative elements and developing those elements into suggestions that are safe, factually grounded, or consistent with user constraints. Existing work mainly helps users detect, verify, or inspect suspicious content \cite{nahar_fakes_2024,cheng_relic_2024,leiser_hill_2024}. Our finding addresses what an LLM could do between detection and removal. This extends hallucination mitigation toward preserving a possible source of inspiration without retaining the erroneous claim.

\subsubsection{Support the Inspired-to State}

\textbf{After users enter \emph{inspired-to}, user-directed \emph{salvage support} should help them develop the inspiration while removing the error.} Existing LLM tools let users select and rewrite or expand local passages \cite{ippolito_creative_2022}; intent annotations and compositional co-creative interfaces also support fine-grained, non-linear revision \cite{gmeiner_intent_2025,cao_compositional_2025}. Distinguished from these local editing designs, our proposal applies when the LLM has not identified the hallucination but the user has recognized both the error and its inspiring elements. In this design, participants proposed marking, deleting, or annotating errors, pinning useful content, and preserving revision traces. The LLM should then revise only marked errors while retaining user-selected content and edit records. This addresses a post-inspiration problem that general editing tools do not explicitly consider: correcting an error without discarding the idea that motivated continued work.

\subsection{Limitations and Future Work}

First, as a qualitative study with mainly young creative professionals, participants' views cannot represent all users. Second, we examined creative tasks in research, design, and media content creation; hallucinations may inspire users in other domains, but generalization beyond these settings remains uncertain. Third, inspiration is personal and situated, and our account relies mainly on participants' retrospective recollections and judgments. The same hallucination may produce different outcomes for users with different knowledge, resources, and goals. 

Future research can proceed in the following three directions. One is to study low-risk settings such as fiction writing and advertising ideation, while continuing to prioritize hallucination detection, reduction, and prevention in high-risk or high-responsibility settings such as medicine and law. A second is to follow the full trajectory after a hallucination appears rather than record only whether the generated output contains an error, and develop the evaluation of hallucination consequences proposed in Section \ref{sd:evalhallu}. A third is to implement our design recommendations in practical settings and test whether they improve how users respond to factual and faithfulness hallucinations and help them develop inspiration. These studies could help HCI distinguish when hallucinations must be prevented from how users might treat those already present in low-risk contexts as inspiring.

\section{Conclusion}


This study examined when LLM hallucinations inspire creative practitioners and how these experiences shape subsequent interaction. Across 41 cases and interviews with 20 creative practitioners in research, design, and social media content creation, hallucinations were not inherently useful. They became inspiring when users found goal-relevant possibilities beyond their existing knowledge. Our framework connects these situated judgments to subsequent behaviors and distinguishes the paths through which factual and faithfulness hallucinations led to inspiration. Inspiration from faithfulness hallucinations changed across creative stages, and hallucination-inspired experiences reshaped which LLM roles users trusted. We therefore argue that hallucination evaluation should assess downstream interaction consequences separately from error identification. Stage-sensitive controls, explicit constraint priorities, differentiated output labels, and salvage support could help LLM preserve inspiring ideas while correcting errors, offering the HCI community guidance for the responsible design of LLMs.

\begin{acks}

\end{acks}

\bibliographystyle{ACM-Reference-Format}
\bibliography{ref}

@article{huang_survey_2025,
    title = {A {Survey} on {Hallucination} in {Large} {Language} {Models}: {Principles}, {Taxonomy}, {Challenges}, and {Open} {Questions}},
    volume = {43},
    issn = {1046-8188},
    shorttitle = {A {Survey} on {Hallucination} in {Large} {Language} {Models}},
    url = {https://dl.acm.org/doi/10.1145/3703155},
    doi = {10.1145/3703155},
    language = {en-US},
    number = {2},
    urldate = {2025-11-19},
    journal = {ACM Trans. Inf. Syst.},
    author = {Huang, Lei and Yu, Weijiang and Ma, Weitao and Zhong, Weihong and Feng, Zhangyin and Wang, Haotian and Chen, Qianglong and Peng, Weihua and Feng, Xiaocheng and Qin, Bing and Liu, Ting},
    year = {2025},
    pages = {42:1--42:55},
}

@misc{rawte_troubling_2023,
    title = {The {Troubling} {Emergence} of {Hallucination} in {Large} {Language} {Models} -- {An} {Extensive} {Definition}, {Quantification}, and {Prescriptive} {Remediations}},
    url = {http://arxiv.org/abs/2310.04988},
    doi = {10.48550/arXiv.2310.04988},
    urldate = {2025-12-04},
    publisher = {arXiv},
    author = {Rawte, Vipula and Chakraborty, Swagata and Pathak, Agnibh and Sarkar, Anubhav and Tonmoy, S. M. Towhidul Islam and Chadha, Aman and Sheth, Amit P. and Das, Amitava},
    month = oct,
    year = {2023},
    note = {arXiv:2310.04988 [cs]},
}

@article{ji_survey_2023,
    title = {Survey of {Hallucination} in {Natural} {Language} {Generation}},
    volume = {55},
    issn = {0360-0300},
    url = {https://dl.acm.org/doi/10.1145/3571730},
    doi = {10.1145/3571730},
    number = {12},
    urldate = {2025-12-01},
    journal = {ACM Comput. Surv.},
    author = {Ji, Ziwei and Lee, Nayeon and Frieske, Rita and Yu, Tiezheng and Su, Dan and Xu, Yan and Ishii, Etsuko and Bang, Ye Jin and Madotto, Andrea and Fung, Pascale},
    year = {2023},
    pages = {248:1--248:38},
}

@misc{ye_cognitive_2023,
    title = {Cognitive {Mirage}: {A} {Review} of {Hallucinations} in {Large} {Language} {Models}},
    shorttitle = {Cognitive {Mirage}},
    url = {https://arxiv.org/abs/2309.06794v1},
    doi = {10.48550/arXiv.2309.06794},
    language = {en},
    urldate = {2025-12-01},
    publisher = {arXiv},
    author = {Ye, Hongbin and Liu, Tong and Zhang, Aijia and Hua, Wei and Jia, Weiqiang},
    month = sep,
    year = {2023},
}

@misc{li_lokis_2025,
    title = {Loki's {Dance} of {Illusions}: {A} {Comprehensive} {Survey} of {Hallucination} in {Large} {Language} {Models}},
    shorttitle = {Loki's {Dance} of {Illusions}},
    url = {http://arxiv.org/abs/2507.02870},
    doi = {10.48550/arXiv.2507.02870},
    urldate = {2025-12-01},
    publisher = {arXiv},
    author = {Li, Chaozhuo and Wang, Pengbo and Wang, Chenxu and Zhang, Litian and Liu, Zheng and Ye, Qiwei and Xu, Yuanbo and Huang, Feiran and Zhang, Xi and Yu, Philip S.},
    month = jun,
    year = {2025},
    note = {arXiv:2507.02870 [cs]},
}

@misc{zhang_sirens_2025,
    title = {Siren's {Song} in the {AI} {Ocean}: {A} {Survey} on {Hallucination} in {Large} {Language} {Models}},
    shorttitle = {Siren's {Song} in the {AI} {Ocean}},
    url = {http://arxiv.org/abs/2309.01219},
    doi = {10.48550/arXiv.2309.01219},
    urldate = {2025-11-19},
    publisher = {arXiv},
    author = {Zhang, Yue and Li, Yafu and Cui, Leyang and Cai, Deng and Liu, Lemao and Fu, Tingchen and Huang, Xinting and Zhao, Enbo and Zhang, Yu and Xu, Chen and Chen, Yulong and Wang, Longyue and Luu, Anh Tuan and Bi, Wei and Shi, Freda and Shi, Shuming},
    month = sep,
    year = {2025},
    note = {arXiv:2309.01219 [cs]},
}

@article{ray_chatgpt_2023,
    title = {{ChatGPT}: {A} comprehensive review on background, applications, key challenges, bias, ethics, limitations and future scope},
    volume = {3},
    issn = {2667-3452},
    shorttitle = {{ChatGPT}},
    url = {https://www.sciencedirect.com/science/article/pii/S266734522300024X},
    doi = {10.1016/j.iotcps.2023.04.003},
    urldate = {2025-12-04},
    journal = {Internet of Things and Cyber-Physical Systems},
    author = {Ray, Partha Pratim},
    month = jan,
    year = {2023},
    pages = {121--154},
}

@inproceedings{mckenna_sources_2023,
    address = {Singapore},
    title = {Sources of {Hallucination} by {Large} {Language} {Models} on {Inference} {Tasks}},
    url = {https://aclanthology.org/2023.findings-emnlp.182},
    doi = {10.18653/v1/2023.findings-emnlp.182},
    language = {en},
    urldate = {2025-11-30},
    booktitle = {Findings of the {Association} for {Computational} {Linguistics}: {EMNLP} 2023},
    publisher = {Association for Computational Linguistics},
    author = {McKenna, Nick and Li, Tianyi and Cheng, Liang and Hosseini, Mohammad and Johnson, Mark and Steedman, Mark},
    year = {2023},
    pages = {2758--2774},
}

@misc{xu_hallucination_2025,
    title = {Hallucination is {Inevitable}: {An} {Innate} {Limitation} of {Large} {Language} {Models}},
    shorttitle = {Hallucination is {Inevitable}},
    url = {http://arxiv.org/abs/2401.11817},
    doi = {10.48550/arXiv.2401.11817},
    urldate = {2025-11-19},
    publisher = {arXiv},
    author = {Xu, Ziwei and Jain, Sanjay and Kankanhalli, Mohan},
    month = feb,
    year = {2025},
    note = {arXiv:2401.11817 [cs]},
}

@article{dahl_large_2024,
    title = {Large {Legal} {Fictions}: {Profiling} {Legal} {Hallucinations} in {Large} {Language} {Models}},
    volume = {16},
    issn = {2161-7201},
    shorttitle = {Large {Legal} {Fictions}},
    url = {https://doi.org/10.1093/jla/laae003},
    doi = {10.1093/jla/laae003},
    number = {1},
    urldate = {2025-11-30},
    journal = {Journal of Legal Analysis},
    author = {Dahl, Matthew and Magesh, Varun and Suzgun, Mirac and Ho, Daniel E},
    month = jan,
    year = {2024},
    pages = {64--93},
}

@inproceedings{pal_med-halt_2023,
    address = {Singapore},
    title = {Med-{HALT}: {Medical} {Domain} {Hallucination} {Test} for {Large} {Language} {Models}},
    shorttitle = {Med-{HALT}},
    url = {https://aclanthology.org/2023.conll-1.21},
    doi = {10.18653/v1/2023.conll-1.21},
    language = {en},
    urldate = {2025-12-01},
    booktitle = {Proceedings of the 27th {Conference} on {Computational} {Natural} {Language} {Learning} ({CoNLL})},
    publisher = {Association for Computational Linguistics},
    author = {Pal, Ankit and Umapathi, Logesh Kumar and Sankarasubbu, Malaikannan},
    year = {2023},
    pages = {314--334},
}

@inproceedings{spatharioti_effects_2025,
    address = {New York, NY, USA},
    series = {{CHI} '25},
    title = {Effects of {LLM}-based {Search} on {Decision} {Making}: {Speed}, {Accuracy}, and {Overreliance}},
    isbn = {979-8-4007-1394-1},
    shorttitle = {Effects of {LLM}-based {Search} on {Decision} {Making}},
    url = {https://dl.acm.org/doi/10.1145/3706598.3714082},
    doi = {10.1145/3706598.3714082},
    urldate = {2025-11-18},
    booktitle = {Proceedings of the 2025 {CHI} {Conference} on {Human} {Factors} in {Computing} {Systems}},
    publisher = {Association for Computing Machinery},
    author = {Spatharioti, Sofia Eleni and Rothschild, David and Goldstein, Daniel G and Hofman, Jake M},
    year = {2025},
    pages = {1--15},
}

@article{lin_towards_2024,
    title = {Towards trustworthy {LLMs}: a review on debiasing and dehallucinating in large language models},
    volume = {57},
    issn = {1573-7462},
    shorttitle = {Towards trustworthy {LLMs}},
    url = {https://doi.org/10.1007/s10462-024-10896-y},
    doi = {10.1007/s10462-024-10896-y},
    language = {en},
    number = {9},
    urldate = {2025-11-30},
    journal = {Artificial Intelligence Review},
    author = {Lin, Zichao and Guan, Shuyan and Zhang, Wending and Zhang, Huiyan and Li, Yugang and Zhang, Huaping},
    month = aug,
    year = {2024},
    pages = {243},
}

@inproceedings{yin_lies_2024,
    address = {New York, NY, USA},
    series = {{CHI} '24},
    title = {Lies, {Deceit}, and {Hallucinations}: {Player} {Perception} and {Expectations} {Regarding} {Trust} and {Deception} in {Games}},
    isbn = {979-8-4007-0330-0},
    shorttitle = {Lies, {Deceit}, and {Hallucinations}},
    url = {https://dl.acm.org/doi/10.1145/3613904.3642253},
    doi = {10.1145/3613904.3642253},
    urldate = {2025-11-18},
    booktitle = {Proceedings of the 2024 {CHI} {Conference} on {Human} {Factors} in {Computing} {Systems}},
    publisher = {Association for Computing Machinery},
    author = {Yin, Michael and Wang, Emi and Ng, Chuoxi and Xiao, Robert},
    year = {2024},
    pages = {1--15},
}

@inproceedings{li_halueval_2023,
    address = {Singapore},
    title = {{HaluEval}: {A} {Large}-{Scale} {Hallucination} {Evaluation} {Benchmark} for {Large} {Language} {Models}},
    shorttitle = {{HaluEval}},
    url = {https://aclanthology.org/2023.emnlp-main.397/},
    doi = {10.18653/v1/2023.emnlp-main.397},
    urldate = {2025-11-30},
    booktitle = {Proceedings of the 2023 {Conference} on {Empirical} {Methods} in {Natural} {Language} {Processing}},
    publisher = {Association for Computational Linguistics},
    author = {Li, Junyi and Cheng, Xiaoxue and Zhao, Xin and Nie, Jian-Yun and Wen, Ji-Rong},
    editor = {Bouamor, Houda and Pino, Juan and Bali, Kalika},
    month = dec,
    year = {2023},
    pages = {6449--6464},
}

@inproceedings{leiser_hill_2024,
    address = {New York, NY, USA},
    series = {{CHI} '24},
    title = {{HILL}: {A} {Hallucination} {Identifier} for {Large} {Language} {Models}},
    isbn = {979-8-4007-0330-0},
    shorttitle = {{HILL}},
    url = {https://dl.acm.org/doi/10.1145/3613904.3642428},
    doi = {10.1145/3613904.3642428},
    urldate = {2025-11-18},
    booktitle = {Proceedings of the 2024 {CHI} {Conference} on {Human} {Factors} in {Computing} {Systems}},
    publisher = {Association for Computing Machinery},
    author = {Leiser, Florian and Eckhardt, Sven and Leuthe, Valentin and Knaeble, Merlin and Mädche, Alexander and Schwabe, Gerhard and Sunyaev, Ali},
    year = {2024},
    pages = {1--13},
}

@article{farquhar_detecting_2024,
    title = {Detecting hallucinations in large language models using semantic entropy},
    volume = {630},
    copyright = {2024 The Author(s)},
    issn = {1476-4687},
    url = {https://www.nature.com/articles/s41586-024-07421-0},
    doi = {10.1038/s41586-024-07421-0},
    language = {en},
    number = {8017},
    urldate = {2025-11-19},
    journal = {Nature},
    author = {Farquhar, Sebastian and Kossen, Jannik and Kuhn, Lorenz and Gal, Yarin},
    month = jun,
    year = {2024},
    note = {Publisher: Nature Publishing Group},
    pages = {625--630},
}

@inproceedings{reinhard_fact_2025,
    address = {New York, NY, USA},
    series = {{CHI} {EA} '25},
    title = {Fact or {Fiction}? {Exploring} {Explanations} to {Identify} {Factual} {Confabulations} in {RAG}-{Based} {LLM} {Systems}},
    isbn = {979-8-4007-1395-8},
    shorttitle = {Fact or {Fiction}?},
    url = {https://dl.acm.org/doi/10.1145/3706599.3720249},
    doi = {10.1145/3706599.3720249},
    urldate = {2025-11-18},
    booktitle = {Proceedings of the {Extended} {Abstracts} of the {CHI} {Conference} on {Human} {Factors} in {Computing} {Systems}},
    publisher = {Association for Computing Machinery},
    author = {Reinhard, Philipp and Li, Mahei Manhai and Fina, Matteo and Leimeister, Jan Marco},
    year = {2025},
    pages = {1--13},
}

@inproceedings{ehsan_new_2025,
    address = {New York, NY, USA},
    series = {{CHI} {EA} '25},
    title = {New {Frontiers} of {Human}-centered {Explainable} {AI} ({HCXAI}): {Participatory} {Civic} {AI}, {Benchmarking} {LLMs}, {XAI} {Hallucinations}, and {Responsible} {AI} {Audits}},
    isbn = {979-8-4007-1395-8},
    shorttitle = {New {Frontiers} of {Human}-centered {Explainable} {AI} ({HCXAI})},
    url = {https://dl.acm.org/doi/10.1145/3706599.3706713},
    doi = {10.1145/3706599.3706713},
    urldate = {2025-11-18},
    booktitle = {Proceedings of the {Extended} {Abstracts} of the {CHI} {Conference} on {Human} {Factors} in {Computing} {Systems}},
    publisher = {Association for Computing Machinery},
    author = {Ehsan, Upol and Watkins, Elizabeth A and Wintersberger, Philipp and Manger, Carina and Hubig, Nina and Savage, Saiph and Weisz, Justin D. and Riener, Andreas},
    year = {2025},
    pages = {1--6},
}

@misc{tonmoy_comprehensive_2024,
    title = {A {Comprehensive} {Survey} of {Hallucination} {Mitigation} {Techniques} in {Large} {Language} {Models}},
    url = {http://arxiv.org/abs/2401.01313},
    doi = {10.48550/arXiv.2401.01313},
    urldate = {2025-12-01},
    publisher = {arXiv},
    author = {Tonmoy, S. M. Towhidul Islam and Zaman, S. M. Mehedi and Jain, Vinija and Rani, Anku and Rawte, Vipula and Chadha, Aman and Das, Amitava},
    month = jan,
    year = {2024},
    note = {arXiv:2401.01313 [cs]},
}

@inproceedings{lewis_retrieval-augmented_2020,
    title = {Retrieval-{Augmented} {Generation} for {Knowledge}-{Intensive} {NLP} {Tasks}},
    volume = {33},
    url = {https://proceedings.neurips.cc/paper/2020/hash/6b493230205f780e1bc26945df7481e5-Abstract.html},
    language = {en-US},
    urldate = {2025-12-04},
    booktitle = {Advances in {Neural} {Information} {Processing} {Systems}},
    publisher = {Curran Associates, Inc.},
    author = {Lewis, Patrick and Perez, Ethan and Piktus, Aleksandra and Petroni, Fabio and Karpukhin, Vladimir and Goyal, Naman and Küttler, Heinrich and Lewis, Mike and Yih, Wen-tau and Rocktäschel, Tim and Riedel, Sebastian and Kiela, Douwe},
    year = {2020},
    pages = {9459--9474},
    address = {New York, NY, USA},
}

@inproceedings{shi_hallucination_2023,
    address = {Singapore},
    title = {Hallucination {Mitigation} in {Natural} {Language} {Generation} from {Large}-{Scale} {Open}-{Domain} {Knowledge} {Graphs}},
    url = {https://aclanthology.org/2023.emnlp-main.770/},
    doi = {10.18653/v1/2023.emnlp-main.770},
    urldate = {2025-12-01},
    booktitle = {Proceedings of the 2023 {Conference} on {Empirical} {Methods} in {Natural} {Language} {Processing}},
    publisher = {Association for Computational Linguistics},
    author = {Shi, Xiao and Zhu, Zhengyuan and Zhang, Zeyu and Li, Chengkai},
    editor = {Bouamor, Houda and Pino, Juan and Bali, Kalika},
    month = dec,
    year = {2023},
    pages = {12506--12521},
}

@inproceedings{agrawal_can_2024,
    address = {Mexico City, Mexico},
    title = {Can {Knowledge} {Graphs} {Reduce} {Hallucinations} in {LLMs}? : {A} {Survey}},
    shorttitle = {Can {Knowledge} {Graphs} {Reduce} {Hallucinations} in {LLMs}?},
    url = {https://aclanthology.org/2024.naacl-long.219},
    doi = {10.18653/v1/2024.naacl-long.219},
    language = {en},
    urldate = {2025-11-30},
    booktitle = {Proceedings of the 2024 {Conference} of the {North} {American} {Chapter} of the {Association} for {Computational} {Linguistics}: {Human} {Language} {Technologies} ({Volume} 1: {Long} {Papers})},
    publisher = {Association for Computational Linguistics},
    author = {Agrawal, Garima and Kumarage, Tharindu and Alghamdi, Zeyad and Liu, Huan},
    year = {2024},
    pages = {3947--3960},
}

@misc{wang_self-consistency_2023,
    title = {Self-{Consistency} {Improves} {Chain} of {Thought} {Reasoning} in {Language} {Models}},
    url = {http://arxiv.org/abs/2203.11171},
    doi = {10.48550/arXiv.2203.11171},
    urldate = {2025-12-04},
    publisher = {arXiv},
    author = {Wang, Xuezhi and Wei, Jason and Schuurmans, Dale and Le, Quoc and Chi, Ed and Narang, Sharan and Chowdhery, Aakanksha and Zhou, Denny},
    month = mar,
    year = {2023},
    note = {arXiv:2203.11171 [cs]},
}

@inproceedings{cheng_relic_2024,
    address = {New York, NY, USA},
    series = {{CHI} '24},
    title = {{RELIC}: {Investigating} {Large} {Language} {Model} {Responses} using {Self}-{Consistency}},
    isbn = {979-8-4007-0330-0},
    shorttitle = {{RELIC}},
    url = {https://dl.acm.org/doi/10.1145/3613904.3641904},
    doi = {10.1145/3613904.3641904},
    urldate = {2025-11-18},
    booktitle = {Proceedings of the 2024 {CHI} {Conference} on {Human} {Factors} in {Computing} {Systems}},
    publisher = {Association for Computing Machinery},
    author = {Cheng, Furui and Zouhar, Vilém and Arora, Simran and Sachan, Mrinmaya and Strobelt, Hendrik and El-Assady, Mennatallah},
    year = {2024},
    pages = {1--18},
}

@inproceedings{dhuliawala_chain--verification_2024,
    address = {Bangkok, Thailand},
    title = {Chain-of-{Verification} {Reduces} {Hallucination} in {Large} {Language} {Models}},
    url = {https://aclanthology.org/2024.findings-acl.212/},
    doi = {10.18653/v1/2024.findings-acl.212},
    urldate = {2025-12-04},
    booktitle = {Findings of the {Association} for {Computational} {Linguistics}: {ACL} 2024},
    publisher = {Association for Computational Linguistics},
    author = {Dhuliawala, Shehzaad and Komeili, Mojtaba and Xu, Jing and Raileanu, Roberta and Li, Xian and Celikyilmaz, Asli and Weston, Jason},
    editor = {Ku, Lun-Wei and Martins, Andre and Srikumar, Vivek},
    month = aug,
    year = {2024},
    pages = {3563--3578},
}

@inproceedings{meng_locating_2022,
    title = {Locating and {Editing} {Factual} {Associations} in {GPT}},
    volume = {35},
    url = {https://proceedings.neurips.cc/paper_files/paper/2022/hash/6f1d43d5a82a37e89b0665b33bf3a182-Abstract-Conference.html},
    language = {en},
    urldate = {2025-12-04},
    booktitle = {Advances in {Neural} {Information} {Processing} {Systems}},
    author = {Meng, Kevin and Bau, David and Andonian, Alex and Belinkov, Yonatan},
    month = dec,
    year = {2022},
    pages = {17359--17372},
    publisher = {Curran Associates, Inc.},
    address = {New York, NY, USA},
}

@misc{meng_mass-editing_2023,
    title = {Mass-{Editing} {Memory} in a {Transformer}},
    url = {http://arxiv.org/abs/2210.07229},
    doi = {10.48550/arXiv.2210.07229},
    urldate = {2025-12-04},
    publisher = {arXiv},
    author = {Meng, Kevin and Sharma, Arnab Sen and Andonian, Alex and Belinkov, Yonatan and Bau, David},
    month = aug,
    year = {2023},
    note = {arXiv:2210.07229 [cs]},
}

@misc{pilcher_purposefully_2025,
    title = {Purposefully {Induced} {Psychosis} ({PIP}): {Embracing} {Hallucination} as {Imagination} in {Large} {Language} {Models}},
    shorttitle = {Purposefully {Induced} {Psychosis} ({PIP})},
    url = {http://arxiv.org/abs/2504.12012},
    doi = {10.48550/arXiv.2504.12012},
    urldate = {2025-11-30},
    publisher = {arXiv},
    author = {Pilcher, Kris and Tütüncü, Esen K.},
    month = apr,
    year = {2025},
    note = {arXiv:2504.12012 [cs]},
}

@article{lee_mathematical_2023,
    title = {A {Mathematical} {Investigation} of {Hallucination} and {Creativity} in {GPT} {Models}},
    volume = {11},
    copyright = {http://creativecommons.org/licenses/by/3.0/},
    issn = {2227-7390},
    url = {https://www.mdpi.com/2227-7390/11/10/2320},
    doi = {10.3390/math11102320},
    language = {en},
    number = {10},
    urldate = {2025-11-19},
    journal = {Mathematics},
    author = {Lee, Minhyeok},
    month = may,
    year = {2023},
    note = {Company: Multidisciplinary Digital Publishing Institute
Distributor: Multidisciplinary Digital Publishing Institute
Institution: Multidisciplinary Digital Publishing Institute
Label: Multidisciplinary Digital Publishing Institute
Publisher: publisher},
    pages = {2320},
}

@misc{he_shakespearean_2025,
    title = {Shakespearean {Sparks}: {The} {Dance} of {Hallucination} and {Creativity} in {LLMs}' {Decoding} {Layers}},
    shorttitle = {Shakespearean {Sparks}},
    url = {http://arxiv.org/abs/2503.02851},
    doi = {10.48550/arXiv.2503.02851},
    urldate = {2025-12-01},
    publisher = {arXiv},
    author = {He, Zicong and Zhang, Boxuan and Cheng, Lu},
    month = mar,
    year = {2025},
    note = {arXiv:2503.02851 [cs]},
}

@misc{jiang_survey_2024,
    title = {A {Survey} on {Large} {Language} {Model} {Hallucination} via a {Creativity} {Perspective}},
    url = {http://arxiv.org/abs/2402.06647},
    doi = {10.48550/arXiv.2402.06647},
    language = {en-US},
    urldate = {2025-11-19},
    publisher = {arXiv},
    author = {Jiang, Xuhui and Tian, Yuxing and Hua, Fengrui and Xu, Chengjin and Wang, Yuanzhuo and Guo, Jian},
    month = feb,
    year = {2024},
    note = {arXiv:2402.06647 [cs]},
}

@article{thrash_inspiration_2003,
    title = {Inspiration as a psychological construct},
    volume = {84},
    issn = {1939-1315},
    doi = {10.1037/0022-3514.84.4.871},
    number = {4},
    journal = {Journal of Personality and Social Psychology},
    author = {Thrash, Todd M. and Elliot, Andrew J.},
    year = {2003},
    note = {Place: US
Publisher: American Psychological Association},
    pages = {871--889},
}

@article{thrash_inspiration_2004,
    title = {Inspiration: core characteristics, component processes, antecedents, and function},
    volume = {87},
    issn = {0022-3514},
    shorttitle = {Inspiration},
    doi = {10.1037/0022-3514.87.6.957},
    language = {eng},
    number = {6},
    journal = {Journal of Personality and Social Psychology},
    author = {Thrash, Todd M. and Elliot, Andrew J.},
    month = dec,
    year = {2004},
    pmid = {15598117},
    pages = {957--973},
}

@article{thrash_psychology_2014,
    title = {The {Psychology} of {Inspiration}},
    volume = {8},
    copyright = {© 2014 John Wiley \& Sons Ltd},
    issn = {1751-9004},
    url = {https://onlinelibrary.wiley.com/doi/abs/10.1111/spc3.12127},
    doi = {10.1111/spc3.12127},
    language = {en},
    number = {9},
    urldate = {2024-03-16},
    journal = {Social and Personality Psychology Compass},
    author = {Thrash, Todd M. and Moldovan, Emil G. and Oleynick, Victoria C. and Maruskin, Laura A.},
    year = {2014},
    note = {\_eprint: https://onlinelibrary.wiley.com/doi/pdf/10.1111/spc3.12127},
    pages = {495--510},
}

@inproceedings{kang_biospark_2024,
    address = {New York, NY, USA},
    series = {{CHI} {EA} '24},
    title = {{BioSpark}: {An} {End}-to-{End} {Generative} {System} for {Biological}-{Analogical} {Inspirations} and {Ideation}},
    isbn = {979-8-4007-0331-7},
    shorttitle = {{BioSpark}},
    url = {https://dl.acm.org/doi/10.1145/3613905.3651035},
    doi = {10.1145/3613905.3651035},
    language = {en-US},
    urldate = {2025-11-18},
    booktitle = {Extended {Abstracts} of the {CHI} {Conference} on {Human} {Factors} in {Computing} {Systems}},
    publisher = {Association for Computing Machinery},
    author = {Kang, Hyeonsu B and Lin, David Chuan-En and Martelaro, Nikolas and Kittur, Aniket and Chen, Yan-Ying and Hong, Matthew K.},
    year = {2024},
    pages = {1--13},
}

@inproceedings{lin_seeking_2025,
    address = {New York, NY, USA},
    series = {{CHI} '25},
    title = {Seeking {Inspiration} through {Human}-{LLM} {Interaction}},
    isbn = {979-8-4007-1394-1},
    url = {https://dl.acm.org/doi/10.1145/3706598.3713259},
    doi = {10.1145/3706598.3713259},
    urldate = {2025-05-28},
    booktitle = {Proceedings of the 2025 {CHI} {Conference} on {Human} {Factors} in {Computing} {Systems}},
    publisher = {Association for Computing Machinery},
    author = {Lin, Xinrui and Huang, Heyan and Huang, Kaihuang and Shu, Xin and Vines, John},
    year = {2025},
    pages = {1--17},
}

@article{bottger_customer_2017,
    title = {Customer {Inspiration}: {Conceptualization}, {Scale} {Development}, and {Validation}},
    volume = {81},
    issn = {0022-2429},
    shorttitle = {Customer {Inspiration}},
    url = {https://doi.org/10.1509/jm.15.0007},
    doi = {10.1509/jm.15.0007},
    language = {en},
    number = {6},
    urldate = {2024-03-06},
    journal = {Journal of Marketing},
    author = {Böttger, Tim and Rudolph, Thomas and Evanschitzky, Heiner and Pfrang, Thilo},
    month = nov,
    year = {2017},
    note = {Publisher: SAGE Publications Inc},
    pages = {116--131},
}

@incollection{thrash_chapter_2021,
    address = {Amsterdam, Netherlands},
    series = {Advances in {Motivation} {Science}},
    title = {Chapter {Five} - {The} creation and curation of all things worthy: {Inspiration} as vital force in persons and cultures},
    volume = {8},
    shorttitle = {Chapter {Five} - {The} creation and curation of all things worthy},
    url = {https://www.sciencedirect.com/science/article/pii/S221509192030002X},
    urldate = {2024-03-06},
    booktitle = {Advances in {Motivation} {Science}},
    publisher = {Elsevier},
    author = {Thrash, Todd M.},
    editor = {Elliot, Andrew J.},
    month = jan,
    year = {2021},
    doi = {10.1016/bs.adms.2020.01.002},
    pages = {181--244},
}

@article{thrash_inspiration_2010,
    title = {Inspiration and the promotion of well-being: {Tests} of causality and mediation},
    volume = {98},
    issn = {1939-1315},
    shorttitle = {Inspiration and the promotion of well-being},
    doi = {10.1037/a0017906},
    number = {3},
    journal = {Journal of Personality and Social Psychology},
    publisher = {American Psychological Association},
    author = {Thrash, Todd M. and Elliot, Andrew J. and Maruskin, Laura A. and Cassidy, Scott E.},
    year = {2010},
    note = {Place: US},
    pages = {488--506},
}

@article{thrash_mediating_2010,
    title = {Mediating between the muse and the masses: {Inspiration} and the actualization of creative ideas},
    volume = {98},
    issn = {1939-1315},
    shorttitle = {Mediating between the muse and the masses},
    doi = {10.1037/a0017907},
    number = {3},
    journal = {Journal of Personality and Social Psychology},
    publisher = {American Psychological Association},
    author = {Thrash, Todd M. and Maruskin, Laura A. and Cassidy, Scott E. and Fryer, James W. and Ryan, Richard M.},
    year = {2010},
    note = {Place: US},
    pages = {469--487},
}

@article{baird_inspired_2012,
    title = {Inspired by {Distraction}: {Mind} {Wandering} {Facilitates} {Creative} {Incubation}},
    volume = {23},
    issn = {0956-7976},
    shorttitle = {Inspired by {Distraction}},
    url = {https://doi.org/10.1177/0956797612446024},
    doi = {10.1177/0956797612446024},
    language = {en},
    number = {10},
    urldate = {2024-09-06},
    journal = {Psychological Science},
    publisher = {SAGE Publications Inc},
    author = {Baird, Benjamin and Smallwood, Jonathan and Mrazek, Michael D. and Kam, Julia W. Y. and Franklin, Michael S. and Schooler, Jonathan W.},
    month = oct,
    year = {2012},
    pages = {1117--1122},
}

@article{stephan_nostalgia-evoked_2015,
    title = {Nostalgia-{Evoked} {Inspiration}: {Mediating} {Mechanisms} and {Motivational} {Implications}},
    volume = {41},
    issn = {0146-1672},
    shorttitle = {Nostalgia-{Evoked} {Inspiration}},
    url = {https://doi.org/10.1177/0146167215596985},
    doi = {10.1177/0146167215596985},
    language = {en},
    number = {10},
    urldate = {2024-12-11},
    journal = {Personality and Social Psychology Bulletin},
    publisher = {SAGE Publications Inc},
    author = {Stephan, Elena and Sedikides, Constantine and Wildschut, Tim and Cheung, Wing-Yee and Routledge, Clay and Arndt, Jamie},
    month = oct,
    year = {2015},
    pages = {1395--1410},
}

@inproceedings{halperin_artificial_2024,
    address = {New York, NY, USA},
    series = {{DIS} '24},
    title = {Artificial {Dreams}: {Surreal} {Visual} {Storytelling} as {Inquiry} {Into} {AI} '{Hallucination}'},
    isbn = {979-8-4007-0583-0},
    shorttitle = {Artificial {Dreams}},
    url = {https://dl.acm.org/doi/10.1145/3643834.3660685},
    doi = {10.1145/3643834.3660685},
    urldate = {2026-05-04},
    booktitle = {Proceedings of the 2024 {ACM} {Designing} {Interactive} {Systems} {Conference}},
    publisher = {Association for Computing Machinery},
    author = {Halperin, Brett A. and Lukin, Stephanie M},
    year = {2024},
    pages = {619--637},
}

@inproceedings{sattele_creative_2025,
    address = {New York, NY, USA},
    series = {C\&amp;{C} '25},
    title = {Creative dialogue: evolving intent in designer-{AI} workflows},
    isbn = {979-8-4007-1289-0},
    shorttitle = {Creative dialogue},
    url = {https://dl.acm.org/doi/10.1145/3698061.3734419},
    doi = {10.1145/3698061.3734419},
    urldate = {2026-05-05},
    booktitle = {Proceedings of the 2025 {Conference} on {Creativity} and {Cognition}},
    publisher = {Association for Computing Machinery},
    author = {Sattele, Vanessa},
    year = {2025},
    pages = {39--44},
}

@inproceedings{pu_ideasynth_2025,
    address = {New York, NY, USA},
    series = {{CHI} '25},
    title = {{IdeaSynth}: {Iterative} {Research} {Idea} {Development} {Through} {Evolving} and {Composing} {Idea} {Facets} with {Literature}-{Grounded} {Feedback}},
    isbn = {979-8-4007-1394-1},
    shorttitle = {{IdeaSynth}},
    url = {https://dl.acm.org/doi/10.1145/3706598.3714057},
    doi = {10.1145/3706598.3714057},
    language = {en-US},
    urldate = {2025-05-28},
    booktitle = {Proceedings of the 2025 {CHI} {Conference} on {Human} {Factors} in {Computing} {Systems}},
    publisher = {Association for Computing Machinery},
    author = {Pu, Kevin and Feng, K. J. Kevin and Grossman, Tovi and Hope, Tom and Dalvi Mishra, Bhavana and Latzke, Matt and Bragg, Jonathan and Chang, Joseph Chee and Siangliulue, Pao},
    year = {2025},
    pages = {1--31},
}

@inproceedings{fan_litlinker_2025,
    address = {New York, NY, USA},
    series = {{CHI} '25},
    title = {{LitLinker}: {Supporting} the {Ideation} of {Interdisciplinary} {Contexts} with {Large} {Language} {Models} for {Teaching} {Literature} in {Elementary} {Schools}},
    isbn = {979-8-4007-1394-1},
    shorttitle = {{LitLinker}},
    url = {https://dl.acm.org/doi/10.1145/3706598.3714111},
    doi = {10.1145/3706598.3714111},
    urldate = {2025-05-27},
    booktitle = {Proceedings of the 2025 {CHI} {Conference} on {Human} {Factors} in {Computing} {Systems}},
    publisher = {Association for Computing Machinery},
    author = {Fan, Haoxiang and Zhou, Changshuang and Yu, Hao and Wu, Xueyang and Gu, Jiangyu and Peng, Zhenhui},
    year = {2025},
    pages = {1--19},
}

@inproceedings{choi_proxona_2025,
    address = {New York, NY, USA},
    series = {{CHI} '25},
    title = {Proxona: {Supporting} {Creators}' {Sensemaking} and {Ideation} with {LLM}-{Powered} {Audience} {Personas}},
    isbn = {979-8-4007-1394-1},
    shorttitle = {Proxona},
    url = {https://dl.acm.org/doi/10.1145/3706598.3714034},
    doi = {10.1145/3706598.3714034},
    urldate = {2025-05-28},
    booktitle = {Proceedings of the 2025 {CHI} {Conference} on {Human} {Factors} in {Computing} {Systems}},
    publisher = {Association for Computing Machinery},
    author = {Choi, Yoonseo and Kang, Eun Jeong and Choi, Seulgi and Lee, Min Kyung and Kim, Juho},
    year = {2025},
    pages = {1--32},
}

@inproceedings{marquardt_imaginationvellum_2025,
    address = {New York, NY, USA},
    series = {{UIST} '25},
    title = {{ImaginationVellum}: {Generative}-{AI} {Ideation} {Canvas} with {Spatial} {Prompts}, {Generative} {Strokes}, and {Ideation} {History}},
    isbn = {979-8-4007-2037-6},
    shorttitle = {{ImaginationVellum}},
    url = {https://dl.acm.org/doi/10.1145/3746059.3747631},
    doi = {10.1145/3746059.3747631},
    urldate = {2026-02-23},
    booktitle = {Proceedings of the 38th {Annual} {ACM} {Symposium} on {User} {Interface} {Software} and {Technology}},
    publisher = {Association for Computing Machinery},
    author = {Marquardt, Nicolai and Roseway, Asta and Romat, Hugo and Panda, Payod and Pahud, Michel and Ramos, Gonzalo and Drucker, Steven M. and Wilson, Andrew D. and Hinckley, Ken and Riche, Nathalie},
    year = {2025},
    pages = {1--19},
}

@inproceedings{tang_how_2026,
    address = {New York, NY, USA},
    series = {{CHI} '26},
    title = {How {Do} {Human} {Creators} {Embrace} {Human}-{AI} {Co}-{Creation}? {A} {Perspective} on {Human} {Agency} of {Screenwriters}},
    isbn = {979-8-4007-2278-3},
    shorttitle = {How {Do} {Human} {Creators} {Embrace} {Human}-{AI} {Co}-{Creation}?},
    url = {https://dl.acm.org/doi/10.1145/3772318.3790300},
    doi = {10.1145/3772318.3790300},
    urldate = {2026-05-04},
    booktitle = {Proceedings of the 2026 {CHI} {Conference} on {Human} {Factors} in {Computing} {Systems}},
    publisher = {Association for Computing Machinery},
    author = {Tang, Yuying and Zhou, Jiayi and Li, Haotian and Xie, Xing and Ma, Xiaojuan and Qu, Huamin},
    year = {2026},
    pages = {1--19},
}

@inproceedings{chakrabarty_creativity_2024,
    address = {New York, NY, USA},
    series = {C\&amp;{C} '24},
    title = {Creativity {Support} in the {Age} of {Large} {Language} {Models}: {An} {Empirical} {Study} {Involving} {Professional} {Writers}},
    isbn = {979-8-4007-0485-7},
    shorttitle = {Creativity {Support} in the {Age} of {Large} {Language} {Models}},
    url = {https://dl.acm.org/doi/10.1145/3635636.3656201},
    doi = {10.1145/3635636.3656201},
    urldate = {2026-05-05},
    booktitle = {Proceedings of the 16th {Conference} on {Creativity} \& {Cognition}},
    publisher = {Association for Computing Machinery},
    author = {Chakrabarty, Tuhin and Padmakumar, Vishakh and Brahman, Faeze and Muresan, Smaranda},
    year = {2024},
    pages = {132--155},
}

@inproceedings{chen_autospark_2024,
    address = {New York, NY, USA},
    series = {{UIST} '24},
    title = {{AutoSpark}: {Supporting} {Automobile} {Appearance} {Design} {Ideation} with {Kansei} {Engineering} and {Generative} {AI}},
    isbn = {979-8-4007-0628-8},
    shorttitle = {{AutoSpark}},
    url = {https://dl.acm.org/doi/10.1145/3654777.3676337},
    doi = {10.1145/3654777.3676337},
    urldate = {2026-02-23},
    booktitle = {Proceedings of the 37th {Annual} {ACM} {Symposium} on {User} {Interface} {Software} and {Technology}},
    publisher = {Association for Computing Machinery},
    author = {Chen, Liuqing and Jing, Qianzhi and Tsang, Yixin and Wang, Qianyi and Liu, Ruocong and Xia, Duowei and Zhou, Yunzhan and Sun, Lingyun},
    year = {2024},
    pages = {1--19},
}

@inproceedings{wang_aideation_2025,
    address = {New York, NY, USA},
    series = {{CHI} '25},
    title = {{AIdeation}: {Designing} a {Human}-{AI} {Collaborative} {Ideation} {System} for {Concept} {Designers}},
    isbn = {979-8-4007-1394-1},
    shorttitle = {{AIdeation}},
    url = {https://dl.acm.org/doi/10.1145/3706598.3714148},
    doi = {10.1145/3706598.3714148},
    urldate = {2025-06-06},
    booktitle = {Proceedings of the 2025 {CHI} {Conference} on {Human} {Factors} in {Computing} {Systems}},
    publisher = {Association for Computing Machinery},
    author = {Wang, Wen-Fan and Lu, Chien-Ting and Ponsa i Campanyà, Nil and Chen, Bing-Yu and Chen, Mike Y.},
    year = {2025},
    pages = {1--28},
}

@misc{argese_hallucination_2026,
    title = {Hallucination or {Creativity}: {How} to {Evaluate} {AI}-{Generated} {Scientific} {Stories}?},
    shorttitle = {Hallucination or {Creativity}},
    url = {http://arxiv.org/abs/2602.02290},
    doi = {10.48550/arXiv.2602.02290},
    urldate = {2026-05-05},
    publisher = {arXiv},
    author = {Argese, Alex and Lisena, Pasquale and Troncy, Raphaël},
    month = mar,
    year = {2026},
    note = {arXiv:2602.02290 [cs]},
}

@article{wan_it_2024,
    title = {"{It} {Felt} {Like} {Having} a {Second} {Mind}": {Investigating} {Human}-{AI} {Co}-creativity in {Prewriting} with {Large} {Language} {Models}},
    volume = {8},
    shorttitle = {"{It} {Felt} {Like} {Having} a {Second} {Mind}"},
    url = {https://dl.acm.org/doi/10.1145/3637361},
    doi = {10.1145/3637361},
    language = {en-US},
    number = {CSCW1},
    urldate = {2026-05-05},
    journal = {Proc. ACM Hum.-Comput. Interact.},
    author = {Wan, Qian and Hu, Siying and Zhang, Yu and Wang, Piaohong and Wen, Bo and Lu, Zhicong},
    year = {2024},
    pages = {84:1--84:26},
}

@inproceedings{bangerl_creaitive_2025,
    address = {New York, NY, USA},
    series = {{CHI} '25},
    title = {{CreAItive} {Collaboration}? {Users}' {Misjudgment} of {AI}-{Creativity} {Affects} {Their} {Collaborative} {Performance}},
    isbn = {979-8-4007-1394-1},
    shorttitle = {{CreAItive} {Collaboration}?},
    url = {https://dl.acm.org/doi/10.1145/3706598.3713886},
    doi = {10.1145/3706598.3713886},
    language = {en-US},
    urldate = {2025-06-06},
    booktitle = {Proceedings of the 2025 {CHI} {Conference} on {Human} {Factors} in {Computing} {Systems}},
    publisher = {Association for Computing Machinery},
    author = {Bangerl, Mia Magdalena and Disch, Leonie and David, Tamara and Pammer-Schindler, Viktoria},
    year = {2025},
    pages = {1--17},
}

@inproceedings{li_beyond_2026,
    address = {New York, NY, USA},
    series = {{CHI} '26},
    title = {Beyond {Input}–{Output}: {Rethinking} {Creativity} through {Design}-by-{Analogy} in {Human}–{AI} {Collaboration}},
    isbn = {979-8-4007-2278-3},
    shorttitle = {Beyond {Input}–{Output}},
    url = {https://dl.acm.org/doi/10.1145/3772318.3791403},
    doi = {10.1145/3772318.3791403},
    urldate = {2026-05-03},
    booktitle = {Proceedings of the 2026 {CHI} {Conference} on {Human} {Factors} in {Computing} {Systems}},
    publisher = {Association for Computing Machinery},
    author = {Li, Xuechen and Zhang, Shuai and Cao, Nan and Chen, Qing},
    year = {2026},
    pages = {1--20},
}

@article{chakraborty_promethean_2024,
    title = {The {Promethean} {Dilemma} of {AI} at the {Intersection} of {Hallucination} and {Creativity}},
    volume = {67},
    issn = {0001-0782},
    url = {https://dl.acm.org/doi/10.1145/3652102},
    doi = {10.1145/3652102},
    number = {10},
    urldate = {2026-05-05},
    journal = {Commun. ACM},
    author = {Chakraborty, Tanmoy and Masud, Sarah},
    year = {2024},
    pages = {26--28},
}

@inproceedings{domanti_are_2026,
    address = {New York, NY, USA},
    series = {{CHI} '26},
    title = {Are {Semantic} {Networks} {Associated} with {Idea} {Originality} in {Artificial} {Creativity}? {A} {Comparison} with {Human} {Agents}},
    isbn = {979-8-4007-2278-3},
    shorttitle = {Are {Semantic} {Networks} {Associated} with {Idea} {Originality} in {Artificial} {Creativity}?},
    url = {https://dl.acm.org/doi/10.1145/3772318.3790849},
    doi = {10.1145/3772318.3790849},
    urldate = {2026-05-04},
    booktitle = {Proceedings of the 2026 {CHI} {Conference} on {Human} {Factors} in {Computing} {Systems}},
    publisher = {Association for Computing Machinery},
    author = {Domanti, Umberto and Campidelli, Lorenzo and Agnoli, Sergio and De Angeli, Antonella},
    year = {2026},
    pages = {1--18},
}

@article{braun_using_2006,
    title = {Using thematic analysis in psychology},
    volume = {3},
    issn = {1478-0887},
    url = {https://www.tandfonline.com/doi/abs/10.1191/1478088706qp063oa},
    doi = {10.1191/1478088706qp063oa},
    number = {2},
    urldate = {2024-03-14},
    journal = {Qualitative Research in Psychology},
    publisher = {Routledge},
    author = {Braun, Virginia and Clarke, Victoria},
    month = jan,
    year = {2006},
    note = {\_eprint: https://www.tandfonline.com/doi/pdf/10.1191/1478088706qp063oa},
    pages = {77--101},
}

@misc{banerjee_does_2026,
    title = {Does {Less} {Hallucination} {Mean} {Less} {Creativity}? {An} {Empirical} {Investigation} in {LLMs}},
    shorttitle = {Does {Less} {Hallucination} {Mean} {Less} {Creativity}?},
    url = {http://arxiv.org/abs/2512.11509},
    doi = {10.48550/arXiv.2512.11509},
    urldate = {2026-05-05},
    publisher = {arXiv},
    author = {Banerjee, Mohor and Wangsajaya, Nadya Yuki and Alsagoff, Syed Ali Redha and Tan, Min Sen and Chun, Zachary Choy Kit and Wei, Alvin Chan Guo},
    month = jan,
    year = {2026},
    note = {arXiv:2512.11509 [cs]},
}

@misc{nahar_fakes_2024,
    title = {Fakes of {Varying} {Shades}: {How} {Warning} {Affects} {Human} {Perception} and {Engagement} {Regarding} {LLM} {Hallucinations}},
    shorttitle = {Fakes of {Varying} {Shades}},
    url = {http://arxiv.org/abs/2404.03745},
    doi = {10.48550/arXiv.2404.03745},
    urldate = {2026-05-04},
    publisher = {arXiv},
    author = {Nahar, Mahjabin and Seo, Haeseung and Lee, Eun-Ju and Xiong, Aiping and Lee, Dongwon},
    month = aug,
    year = {2024},
    note = {arXiv:2404.03745 [cs]},
}

@inproceedings{ashktorab_emerging_2025,
    address = {New York, NY, USA},
    series = {{CHIWORK} '25},
    title = {Emerging {Reliance} {Behaviors} in {Human}-{AI} {Content} {Grounded} {Data} {Generation}: {The} {Role} of {Cognitive} {Forcing} {Functions} and {Hallucinations}},
    isbn = {979-8-4007-1384-2},
    shorttitle = {Emerging {Reliance} {Behaviors} in {Human}-{AI} {Content} {Grounded} {Data} {Generation}},
    url = {https://dl.acm.org/doi/10.1145/3729176.3729179},
    doi = {10.1145/3729176.3729179},
    urldate = {2026-07-16},
    booktitle = {Proceedings of the 4th {Annual} {Symposium} on {Human}-{Computer} {Interaction} for {Work}},
    publisher = {Association for Computing Machinery},
    author = {Ashktorab, Zahra and Desmond, Michael and Pan, Qian and Johnson, James and Brachman, Michelle and Dugan, Casey and Danilevsky, Marina and Geyer, Werner},
    year = {2025},
    pages = {1--17},
}

@inproceedings{liu_we_2024,
    address = {New York, NY, USA},
    series = {{CHI} {EA} '24},
    title = {"{We} {Need} {Structured} {Output}": {Towards} {User}-centered {Constraints} on {Large} {Language} {Model} {Output}},
    isbn = {979-8-4007-0331-7},
    shorttitle = {"{We} {Need} {Structured} {Output}"},
    url = {https://dl.acm.org/doi/10.1145/3613905.3650756},
    doi = {10.1145/3613905.3650756},
    urldate = {2025-11-18},
    booktitle = {Extended {Abstracts} of the {CHI} {Conference} on {Human} {Factors} in {Computing} {Systems}},
    publisher = {Association for Computing Machinery},
    author = {Liu, Michael Xieyang and Liu, Frederick and Fiannaca, Alexander J. and Koo, Terry and Dixon, Lucas and Terry, Michael and Cai, Carrie J.},
    year = {2024},
    pages = {1--9},
}

@article{biskjaer_how_2020,
    title = {How task constraints affect inspiration search strategies},
    volume = {30},
    issn = {1573-1804},
    url = {https://doi.org/10.1007/s10798-019-09496-7},
    doi = {10.1007/s10798-019-09496-7},
    language = {en},
    number = {1},
    urldate = {2024-08-24},
    journal = {International Journal of Technology and Design Education},
    author = {Biskjaer, Michael Mose and Christensen, Bo T. and Friis-Olivarius, Morten and Abildgaard, Sille J. J. and Lundqvist, Caroline and Halskov, Kim},
    month = mar,
    year = {2020},
    pages = {101--125},
}

@misc{ryser_calibrated_2025,
    title = {Calibrated {Trust} in {Dealing} with {LLM} {Hallucinations}: {A} {Qualitative} {Study}},
    shorttitle = {Calibrated {Trust} in {Dealing} with {LLM} {Hallucinations}},
    url = {http://arxiv.org/abs/2512.09088},
    doi = {10.48550/arXiv.2512.09088},
    urldate = {2026-07-16},
    publisher = {arXiv},
    author = {Ryser, Adrian and Allwein, Florian and Schlippe, Tim},
    month = dec,
    year = {2025},
    note = {arXiv:2512.09088 [cs.AI]},
}

@inproceedings{bang_hallulens_2025,
    address = {Vienna, Austria},
    title = {{HalluLens}: {LLM} {Hallucination} {Benchmark}},
    isbn = {979-8-89176-251-0},
    shorttitle = {{HalluLens}},
    url = {https://aclanthology.org/2025.acl-long.1176/},
    doi = {10.18653/v1/2025.acl-long.1176},
    urldate = {2026-07-16},
    booktitle = {Proceedings of the 63rd {Annual} {Meeting} of the {Association} for {Computational} {Linguistics} ({Volume} 1: {Long} {Papers})},
    publisher = {Association for Computational Linguistics},
    author = {Bang, Yejin and Ji, Ziwei and Schelten, Alan and Hartshorn, Anthony and Fowler, Tara and Zhang, Cheng and Cancedda, Nicola and Fung, Pascale},
    editor = {Che, Wanxiang and Nabende, Joyce and Shutova, Ekaterina and Pilehvar, Mohammad Taher},
    month = jul,
    year = {2025},
    pages = {24128--24156},
}

@misc{ren_detecting_2026,
    title = {Detecting {Hallucinations} in {Authentic} {LLM}-{Human} {Interactions}},
    url = {http://arxiv.org/abs/2510.10539},
    doi = {10.48550/arXiv.2510.10539},
    urldate = {2026-07-16},
    publisher = {arXiv},
    author = {Ren, Yujie and Gruhlke, Niklas and Lauscher, Anne},
    month = mar,
    year = {2026},
    note = {arXiv:2510.10539 [cs.CL]},
}

@inproceedings{steinbach_when_2025,
    address = {New York, NY, USA},
    series = {L@{S} '25},
    title = {When {LLMs} {Hallucinate}: {Examining} the {Effects} of {Erroneous} {Feedback} in {Math} {Tutoring} {Systems}},
    isbn = {979-8-4007-1291-3},
    shorttitle = {When {LLMs} {Hallucinate}},
    url = {https://dl.acm.org/doi/10.1145/3698205.3729555},
    doi = {10.1145/3698205.3729555},
    urldate = {2026-07-16},
    booktitle = {Proceedings of the {Twelfth} {ACM} {Conference} on {Learning} @ {Scale}},
    publisher = {Association for Computing Machinery},
    author = {Steinbach, Marlene and Bhandari, Shreya and Meyer, Jennifer and Pardos, Zachary A.},
    year = {2025},
    pages = {139--150},
}

@article{asgari_framework_2025,
    title = {A framework to assess clinical safety and hallucination rates of {LLMs} for medical text summarisation},
    volume = {8},
    copyright = {2025 The Author(s)},
    issn = {2398-6352},
    url = {https://www.nature.com/articles/s41746-025-01670-7},
    doi = {10.1038/s41746-025-01670-7},
    language = {en},
    number = {1},
    urldate = {2026-07-16},
    journal = {npj Digital Medicine},
    publisher = {Nature Publishing Group},
    author = {Asgari, Elham and Montaña-Brown, Nina and Dubois, Magda and Khalil, Saleh and Balloch, Jasmine and Yeung, Joshua Au and Pimenta, Dominic},
    month = may,
    year = {2025},
    pages = {274},
}

@misc{karimi_evaluating_2018,
    title = {Evaluating {Creativity} in {Computational} {Co}-{Creative} {Systems}},
    url = {http://arxiv.org/abs/1807.09886},
    doi = {10.48550/arXiv.1807.09886},
    urldate = {2026-07-16},
    publisher = {arXiv},
    author = {Karimi, Pegah and Grace, Kazjon and Maher, Mary Lou and Davis, Nicholas},
    month = jul,
    year = {2018},
    note = {arXiv:1807.09886 [cs.AI]},
}

@article{bellemare-pepin_divergent_2026,
    title = {Divergent creativity in humans and large language models},
    volume = {16},
    copyright = {2026 The Author(s)},
    issn = {2045-2322},
    url = {https://www.nature.com/articles/s41598-025-25157-3},
    doi = {10.1038/s41598-025-25157-3},
    language = {en},
    number = {1},
    urldate = {2026-07-16},
    journal = {Scientific Reports},
    publisher = {Nature Publishing Group},
    author = {Bellemare-Pepin, Antoine and Lespinasse, François and Thölke, Philipp and Harel, Yann and Mathewson, Kory and Olson, Jay A. and Bengio, Yoshua and Jerbi, Karim},
    month = jan,
    year = {2026},
    pages = {1279},
}

@inproceedings{almeda_prompting_2024,
    address = {New York, NY, USA},
    series = {{CHI} '24},
    title = {Prompting for {Discovery}: {Flexible} {Sense}-{Making} for {AI} {Art}-{Making} with {Dreamsheets}},
    isbn = {979-8-4007-0330-0},
    shorttitle = {Prompting for {Discovery}},
    url = {https://dl.acm.org/doi/10.1145/3613904.3642858},
    doi = {10.1145/3613904.3642858},
    urldate = {2026-05-05},
    booktitle = {Proceedings of the 2024 {CHI} {Conference} on {Human} {Factors} in {Computing} {Systems}},
    publisher = {Association for Computing Machinery},
    author = {Almeda, Shm Garanganao and Zamfirescu-Pereira, J.D. and Kim, Kyu Won and Mani Rathnam, Pradeep and Hartmann, Bjoern},
    year = {2024},
    pages = {1--17},
}

@inproceedings{chung_talebrush_2022,
    address = {New York, NY, USA},
    series = {{CHI} '22},
    title = {{TaleBrush}: {Sketching} {Stories} with {Generative} {Pretrained} {Language} {Models}},
    isbn = {978-1-4503-9157-3},
    shorttitle = {{TaleBrush}},
    url = {https://dl.acm.org/doi/10.1145/3491102.3501819},
    doi = {10.1145/3491102.3501819},
    urldate = {2024-05-03},
    booktitle = {Proceedings of the 2022 {CHI} {Conference} on {Human} {Factors} in {Computing} {Systems}},
    publisher = {Association for Computing Machinery},
    author = {Chung, John Joon Young and Kim, Wooseok and Yoo, Kang Min and Lee, Hwaran and Adar, Eytan and Chang, Minsuk},
    year = {2022},
    pages = {1--19},
}

@misc{rosenbaum_scaffolding_2025,
    title = {Scaffolding {Creativity}: {How} {Divergent} and {Convergent} {LLM} {Personas} {Shape} {Human} {Machine} {Creative} {Problem}-{Solving}},
    shorttitle = {Scaffolding {Creativity}},
    url = {http://arxiv.org/abs/2510.26490},
    doi = {10.48550/arXiv.2510.26490},
    urldate = {2026-07-16},
    publisher = {arXiv},
    author = {Rosenbaum, Alon and David, Yigal and Kaufman, Eran and Ravid, Gilad and Ronen, Amit and Krebs, Assaf},
    month = oct,
    year = {2025},
    note = {arXiv:2510.26490 [cs.HC]},
}

@misc{ippolito_creative_2022,
    title = {Creative {Writing} with an {AI}-{Powered} {Writing} {Assistant}: {Perspectives} from {Professional} {Writers}},
    shorttitle = {Creative {Writing} with an {AI}-{Powered} {Writing} {Assistant}},
    url = {http://arxiv.org/abs/2211.05030},
    doi = {10.48550/arXiv.2211.05030},
    urldate = {2026-07-16},
    publisher = {arXiv},
    author = {Ippolito, Daphne and Yuan, Ann and Coenen, Andy and Burnam, Sehmon},
    month = nov,
    year = {2022},
    note = {arXiv:2211.05030 [cs.HC]},
}

@inproceedings{gmeiner_intent_2025,
    address = {New York, NY, USA},
    series = {{CHI} '25},
    title = {Intent {Tagging}: {Exploring} {Micro}-{Prompting} {Interactions} for {Supporting} {Granular} {Human}-{GenAI} {Co}-{Creation} {Workflows}},
    isbn = {979-8-4007-1394-1},
    shorttitle = {Intent {Tagging}},
    url = {https://dl.acm.org/doi/10.1145/3706598.3713861},
    doi = {10.1145/3706598.3713861},
    urldate = {2025-06-06},
    booktitle = {Proceedings of the 2025 {CHI} {Conference} on {Human} {Factors} in {Computing} {Systems}},
    publisher = {Association for Computing Machinery},
    author = {Gmeiner, Frederic and Marquardt, Nicolai and Bentley, Michael and Romat, Hugo and Pahud, Michel and Brown, David and Roseway, Asta and Martelaro, Nikolas and Holstein, Kenneth and Hinckley, Ken and Riche, Nathalie},
    year = {2025},
    pages = {1--31},
}

@inproceedings{cao_compositional_2025,
    address = {New York, NY, USA},
    series = {{CHI} '25},
    title = {Compositional {Structures} as {Substrates} for {Human}-{AI} {Co}-creation {Environment}: {A} {Design} {Approach} and {A} {Case} {Study}},
    isbn = {979-8-4007-1394-1},
    shorttitle = {Compositional {Structures} as {Substrates} for {Human}-{AI} {Co}-creation {Environment}},
    url = {https://dl.acm.org/doi/10.1145/3706598.3713401},
    doi = {10.1145/3706598.3713401},
    urldate = {2025-11-09},
    booktitle = {Proceedings of the 2025 {CHI} {Conference} on {Human} {Factors} in {Computing} {Systems}},
    publisher = {Association for Computing Machinery},
    author = {Cao, Yining and Huang, Yiyi and Truong, Anh and Shin, Hijung Valentina and Xia, Haijun},
    year = {2025},
    pages = {1--25},
}

@inproceedings{kim_fostering_2025,
    address = {New York, NY, USA},
    series = {{CHI} '25},
    title = {Fostering {Appropriate} {Reliance} on {Large} {Language} {Models}: {The} {Role} of {Explanations}, {Sources}, and {Inconsistencies}},
    isbn = {979-8-4007-1394-1},
    shorttitle = {Fostering {Appropriate} {Reliance} on {Large} {Language} {Models}},
    url = {https://dl.acm.org/doi/10.1145/3706598.3714020},
    doi = {10.1145/3706598.3714020},
    urldate = {2026-07-20},
    booktitle = {Proceedings of the 2025 {CHI} {Conference} on {Human} {Factors} in {Computing} {Systems}},
    publisher = {Association for Computing Machinery},
    author = {Kim, Sunnie S. Y. and Vaughan, Jennifer Wortman and Liao, Q. Vera and Lombrozo, Tania and Russakovsky, Olga},
    year = {2025},
    pages = {1--19},
}

@article{rapp_how_2025,
    title = {How do people react to {ChatGPT}'s unpredictable behavior? {Anthropomorphism}, uncanniness, and fear of {AI}: {A} qualitative study on individuals’ perceptions and understandings of {LLMs}’ nonsensical hallucinations},
    volume = {198},
    issn = {1071-5819},
    shorttitle = {How do people react to {ChatGPT}'s unpredictable behavior?},
    url = {https://www.sciencedirect.com/science/article/pii/S107158192500028X},
    doi = {10.1016/j.ijhcs.2025.103471},
    urldate = {2026-07-20},
    journal = {International Journal of Human-Computer Studies},
    author = {Rapp, Amon and Di Lodovico, Chiara and Di Caro, Luigi},
    month = apr,
    year = {2025},
    pages = {103471},
}

@inproceedings{tian_when_2026,
    address = {New York, NY, USA},
    series = {{CHI} '26},
    title = {When {AI} {Gets} {It} {Wrong}: {Scaffolding} {AI} {Hallucination} {Detection} for {Children} {Through} {Chatbot} {Creation}},
    isbn = {979-8-4007-2278-3},
    shorttitle = {When {AI} {Gets} {It} {Wrong}},
    url = {https://dl.acm.org/doi/10.1145/3772318.3791480},
    doi = {10.1145/3772318.3791480},
    urldate = {2026-05-04},
    booktitle = {Proceedings of the 2026 {CHI} {Conference} on {Human} {Factors} in {Computing} {Systems}},
    publisher = {Association for Computing Machinery},
    author = {Tian, Xiaoyi and Ozturk, Deniz and Edula, Sreekar and Adil, Jibran and Jin, Qiao and Shi, Yang and Barnes, Tiffany},
    year = {2026},
    pages = {1--20},
}

@article{holmstrom_what_2026,
    title = {What is the {Creative} {Value} of {AI} {Hallucinations}? {Insights} from {Generative} {AI} use in {Educational} {Contexts}},
    volume = {28},
    issn = {1572-9419},
    shorttitle = {What is the {Creative} {Value} of {AI} {Hallucinations}?},
    url = {https://doi.org/10.1007/s10796-026-10747-x},
    doi = {10.1007/s10796-026-10747-x},
    language = {en},
    number = {4},
    urldate = {2026-09-03},
    journal = {Information Systems Frontiers},
    author = {Holmström, Jonny and Carroll, Noel and Sundberg, Leif},
    month = aug,
    year = {2026},
    pages = {1389--1406},
}

@misc{lan_schizosyntax_2026,
    title = {Schizo-syntax and the paranoiac guardrail: machinic schizoanalysis of {AI} hallucination and the conditions of a line of flight},
    copyright = {2026 The Author(s)},
    shorttitle = {Schizo-syntax and the paranoiac guardrail},
    url = {https://www.nature.com/articles/s41599-026-08814-4},
    doi = {10.1057/s41599-026-08814-4},
    language = {en},
    urldate = {2026-09-01},
    publisher = {Palgrave},
    author = {Lan, Zihang and Hashim, Azirah and Abdu Saeed, Murad},
    month = aug,
    year = {2026},
}

\appendix

\end{document}